\documentclass{aa}
\usepackage{graphicx}
\usepackage{amsmath}
\usepackage[varg]{txfonts}
\usepackage{natbib,twoopt}
\bibpunct{(}{)}{;}{a}{}{,} 
\makeatletter
\newcommandtwoopt{\citeads}[3][][]{\href{http://adsabs.harvard.edu/abs/#3}%
{\def\hyper@linkstart##1##2{}%
\let\hyper@linkend\@empty\citealp[#1][#2]{#3}}}
\newcommandtwoopt{\citepads}[3][][]{\href{http://adsabs.harvard.edu/abs/#3}%
{\def\hyper@linkstart##1##2{}%
\let\hyper@linkend\@empty\citep[#1][#2]{#3}}}
\newcommandtwoopt{\citetads}[3][][]{\href{http://adsabs.harvard.edu/abs/#3}%
{\def\hyper@linkstart##1##2{}%
\let\hyper@linkend\@empty\citet[#1][#2]{#3}}}
\newcommandtwoopt{\citeyearads}[3][][]%
{\href{http://adsabs.harvard.edu/abs/#3}
{\def\hyper@linkstart##1##2{}%
\let\hyper@linkend\@empty\citeyear[#1][#2]{#3}}}
\makeatother
\begin{document}
   \title{Calibration of \textit{AGILE-GRID} with In-Flight Data and Monte Carlo Simulations}

   \subtitle{}

\author{A.~W.~Chen\inst{\ref{a},\ref{k}}\and 
A.~Argan\inst{\ref{b}}\and
A.~Bulgarelli\inst{\ref{c}}\and
P.~W.~Cattaneo\inst{\ref{d}}\and
T.~Contessi\inst{\ref{a}}\and
A.~Giuliani\inst{\ref{a}}\and
C.~Pittori\inst{\ref{e},\ref{j}}\and
G.~Pucella\inst{\ref{f}}\and
M.~Tavani\inst{\ref{b},\ref{g},\ref{t}}\and
A.~Trois\inst{\ref{h}}\and
F.~Verrecchia\inst{\ref{e},\ref{j}}\and
G.~Barbiellini\inst{\ref{i},\ref{t}}\and
P.~Caraveo\inst{\ref{a}}\and
S.~Colafrancesco\inst{\ref{j},\ref{k}}\and
E.~Costa\inst{\ref{b}}\and
G.~De Paris\inst{\ref{b}}\and
E.~Del Monte\inst{\ref{b}}\and
G.~Di Cocco\inst{\ref{c}}\and
I.~Donnarumma\inst{\ref{b}}\and
Y.~Evangelista\inst{\ref{b}}\and
A.~Ferrari\inst{\ref{l},\ref{t}}\and 
M.~Feroci\inst{\ref{b}}\and
V.~Fioretti\inst{\ref{c}}\and
M.~Fiorini\inst{\ref{a}}\and
F.~Fuschino\inst{\ref{c}}\and
M.~Galli\inst{\ref{m}}\and
F.~Gianotti\inst{\ref{c}}\and
P.~Giommi\inst{\ref{e},\ref{q}}\and
M.~Giusti\inst{\ref{b}}\and
C.~Labanti\inst{\ref{c}}\and 
I.~Lapshov\inst{\ref{b}}\and 
F.~Lazzarotto\inst{\ref{b}}\and
P.~Lipari\inst{\ref{n}}\and
F.~Longo\inst{\ref{i}}\and
F.~Lucarelli\inst{\ref{e},\ref{j}}\and
M.~Marisaldi\inst{\ref{c}}\and
S.~Mereghetti\inst{\ref{a}}\and 
E.~Morelli\inst{\ref{c}}\and 
E.~Moretti\inst{\ref{u},\ref{v}}\and
A.~Morselli\inst{\ref{o}}\and
L.~Pacciani\inst{\ref{b}}\and
A.~Pellizzoni\inst{\ref{h}}\and
F.~Perotti\inst{\ref{a}}\and
G.~Piano\inst{\ref{b},\ref{o},\ref{t}}\and 
P.~Picozza\inst{\ref{g},\ref{o}}\and
M.~Pilia\inst{\ref{h},\ref{s}}\and
M.~Prest\inst{\ref{p}}\and
M.~Rapisarda\inst{\ref{f}}\and
A.~Rappoldi\inst{\ref{d}}\and
A.~Rubini\inst{\ref{b}}\and
S.~Sabatini\inst{\ref{b}}\and
P.~Santolamazza\inst{\ref{e},\ref{j}}\and
P.~Soffitta\inst{\ref{b}}\and
E.~Striani\inst{\ref{b}}\and
M.~Trifoglio\inst{\ref{c}}\and  
G.~Valentini\inst{\ref{q}}\and 
E.~Vallazza\inst{\ref{i}}\and
S.~Vercellone\inst{\ref{r}}\and
V.~Vittorini\inst{\ref{b},\ref{g}}\and
D.~Zanello\inst{\ref{n}}}

\institute{
\label{a}INAF/IASF-Milano, Via E. Bassini, 15 I-20133 Milano, Italy \and 
\label{b}INAF/IAPS, Via Fosso del Cavaliere, 100 I-00133 Roma, Italy \and
\label{c}INAF/IASF-Bologna, Via Gobetti 101, I-40129 Bologna, Italy \and
\label{d}INFN-Pavia, Via Agostino Bassi, 6, I-27100 Pavia, Italy \and
\label{e}ASI Science Data Center, Via Galileo Galilei, I-00044 Frascati (Roma), Italy \and
\label{j}INAF-OAR, Via di Frascati, 33 I-00040, Monteporzio Catone (Roma), Italy \and
\label{f}ENEA Frascati,  Via Enrico Fermi, 13 I-00044 Frascati (Roma), Italy \and
\label{g}Dip. di Fisica, Univ. Tor Vergata, Via della Ricerca Scientifica, 1 I-00133 Roma, Italy \and
\label{t}CIFS, Villa Gualino - v.le Settimio Severo 63, I-10133 Torino, Italy \and
\label{h}INAF-OAC, localita' Poggio dei Pini, strada 54, I-09012 Capoterra (CA), Italy \and
\label{i}Dip. Fisica, Univ. Trieste and INFN Trieste, Via A. Valerio, 2, I-34127 Trieste, Italy \and
\label{k}School of Physics, University of the Witwatersrand, Johannesburg Wits 2050, South Africa \and
\label{l}Dip. Fisica, Universit\'a di Torino, Via Giuria, 1, I-10125, Torino, Italy \and
\label{m}ENEA-Bologna, Via Martiri Montesole, 4 I-40129 Bologna, Italy \and
\label{n}INFN-Roma La Sapienza, P.le A. Moro, 2 I-00185 Roma, Italy \and
\label{o}INFN Roma Tor Vergata, Via della Ricerca Scientifica, 1 I-00133 Roma, Italy \and
\label{p}Dip. di Fisica, Univ. Dell'Insubria, Via Valleggio 11, I-22100 Como, Italy \and
\label{q}Agenzia Spaziale Italiana, Viale Liegi, 26 I-00198 Roma, Italy \and
\label{r}INAF-IASF Palermo, Via Ugo La Malfa 153, I-90146 Palermo, Italy \and
\label{s}ASTRON, the Netherlands Institute for Radio Astronomy, 
Postbus 2, 7990 AA, Dwingeloo, The Netherlands \and
\label{u}Royal Institute of Technology (KTH), Stockholm, Sweden \and
\label{v}The Oskar Klein Centre for Cosmoparticle Physics, Stockholm,Sweden
}

   \date{Received ; accepted }


  \abstract
   {AGILE is a $\gamma$-ray astrophysics mission which has been in orbit since 23 April 2007 and continues to operate reliably. The $\gamma$-ray detector, AGILE-GRID, has
observed Galactic and extragalactic sources, many of which were collected in the
first AGILE Catalog.}
   {We present the calibration of the AGILE-GRID using in-flight data and Monte
   Carlo simulations, producing Instrument Response Functions (IRFs) for the effective area ($A_\mathrm{eff}$), Energy Dispersion Probability (EDP), and Point Spread Function (PSF), each as a function of incident direction in instrument coordinates and energy.}
   {We performed Monte Carlo simulations at different $\gamma$-ray energies and incident angles, including background rejection filters and Kalman filter-based $\gamma$-ray reconstruction.  Long integrations of in-flight observations of the \object{Vela}, \object{Crab} and \object{Geminga} sources in broad and narrow energy bands were used to validate and improve the accuracy of the instrument response functions.}
   {The weighted average PSFs as a function of spectra correspond well to the data for all sources and energy bands.}
   {Changes in the interpolation of the PSF from Monte Carlo data and in the procedure for construction of the energy-weighted effective areas have improved the correspondence between predicted and observed fluxes and spectra of celestial calibration sources, reducing false positives and obviating the need for post-hoc energy-dependent scaling factors. The new IRFs have been publicly available from  the Agile Science Data Centre since November 25, 2011, while the changes in the analysis software will be distributed in an upcoming release.}


   \maketitle
%

\section{Introduction}

AGILE \citepads{2009A&A...502..995T} is an Italian Space Agency (ASI) Small Scientific Mission for
high-energy astrophysics launched on April 23, 2007, composed of a
pair-production Gamma Ray Imager (GRID) sensitive in the energy range 30~MeV-50~GeV (\citeads{2002NIMPA.490..146B}; \citeads{2003NIMPA.501..280P}), an
X-ray Imager (Super-AGILE) sensitive in the energy range 18-60~keV \citepads{2007NIMPA.581..728F}, and a Mini-Calorimeter
sensitive to $\gamma$-rays and charged particles with energies between 300~keV and 100~MeV \citepads{2009NIMPA.598..470L}.
AGILE has
detected both persistent and variable sources, many of which were collected in the
first AGILE Catalog \citepads{2009A&A...506.1563P} and in a recent study of bright sources variability
 (Verrecchia et al. A\&A in press).


\section{Pre-flight calibration of on-board trigger}

The AGILE-GRID is a pair-production telescope with 12 planes of silicon strip detectors, the first 10 of which lie under a pair-conversion tungsten
layer \citepads{2010NIMPA.614..213B}. The size of the tungsten-silicon tracker is $38.06 \times 38.06 \times 21.078 $ cm$^{3}$
and its on-axis depth totals 0.8 radiation lengths.
Monte Carlo simulations (\citeads{2002NIMPA.486..623C}; \citeads{2002NIMPA.486..610L}) with
GEANT3 \citep{GEANT3} were used to determine which
on-board filter strategy would produce the reduction in particle and
albedo background required by telemetry constraints while maintaining an acceptable
effective area for $\gamma$-rays, resulting in hardware on-board triggers \citep{argan}
and on-board simplified Kalman filter \citepads{2006NIMPA.568..692G} for event reconstruction and albedo rejection. These simulations were validated with pre-flight tests with
cosmic-ray muons in the clean rooms of Laben (Milan) and CGS (Tortona)
\citepads{argan} and with $\gamma$-rays at INFN Laboratori Nazionali di Frascati \citepads{2011NIMPA.630..251C, 2012NIMPA.674...55C}.

\section{On-ground background rejection filter}

The effective area ($A_\mathrm{eff}$), the three-dimensional Point Spread Function (PSF), and the Energy Dispersion Probability (EDP) of AGILE-GRID, collectively referred to as the instrument response functions (IRFs), depend on the direction of the incoming $\gamma$-ray in instrument coordinates. Throughout this paper, we will refer to this direction by the angular coordinate $\Omega=(\Theta,\Phi)$, where 
$\Theta$ is the off-axis (polar) angle and $\Phi$ the azimuth angle in spherical coordinates \citepads[see also][]{2002NIMPA.488..295P}.

\subsection{Description}

Additional processing is required on-ground in order to further reduce the particle
background. Detailed analysis of event morphology is used to distinguish $\gamma$-rays from
charged particles. The first on-ground filter to be used with real flight data \emph{F4}, used a hard decision tree and severe cuts for $\gamma$-rays with $\Theta > 40^{\circ}$ to limit contamination by cosmic-ray electrons and positrons. Since AGILE Public Data Release v2.0 in October 6, 2009, \emph{F4} has been replaced by two new filters. A more permissive filter using multi-variate analysis, \emph{FT3ab}, was developed. Further development of the multi-variate analysis technique combined
with some of the \emph{F4} criteria produced a more advanced filter, \emph{FM3.119} (also known as \emph{FM}), which provides a good tradeoff between effective area and background rejection (Bulgarelli et al., in prep.).
Each event is classified as a likely gamma-ray (\emph{G}), uncertain (\emph{L}),
a particle (\emph{P}) or a single-track event (\emph{S}). In practice, all scientific analyses other than pulsar timing and gamma-ray bursts have used \emph{G} events exclusively.

\subsection{Monte Carlo simulations}
\label{sec:MC}

In order to improve and extend the IRFs, we performed additional
Monte Carlo simulations after the launch of AGILE.
For each set of instrument coordinates ($\Theta=
1,\, 30,\, 35,\, 40,\, 45,\, 50,\, 60^{\circ}$
and $\Phi=0,\, 45^{\circ}$), $C_\mathrm{tot}=59\times 10^6$ events
were generated from a source with a power-law spectrum whose
spectral index is $\alpha = -1.7$, with energies ranging from 4~MeV to 50~GeV.
The events were processed using both the on-board filter
and the on-ground event reconstruction procedures, including the background rejection filters.

\subsection{Effective area}
\label{sec:Aeff}

\begin{figure}
    \resizebox{\hsize}{!}{\includegraphics{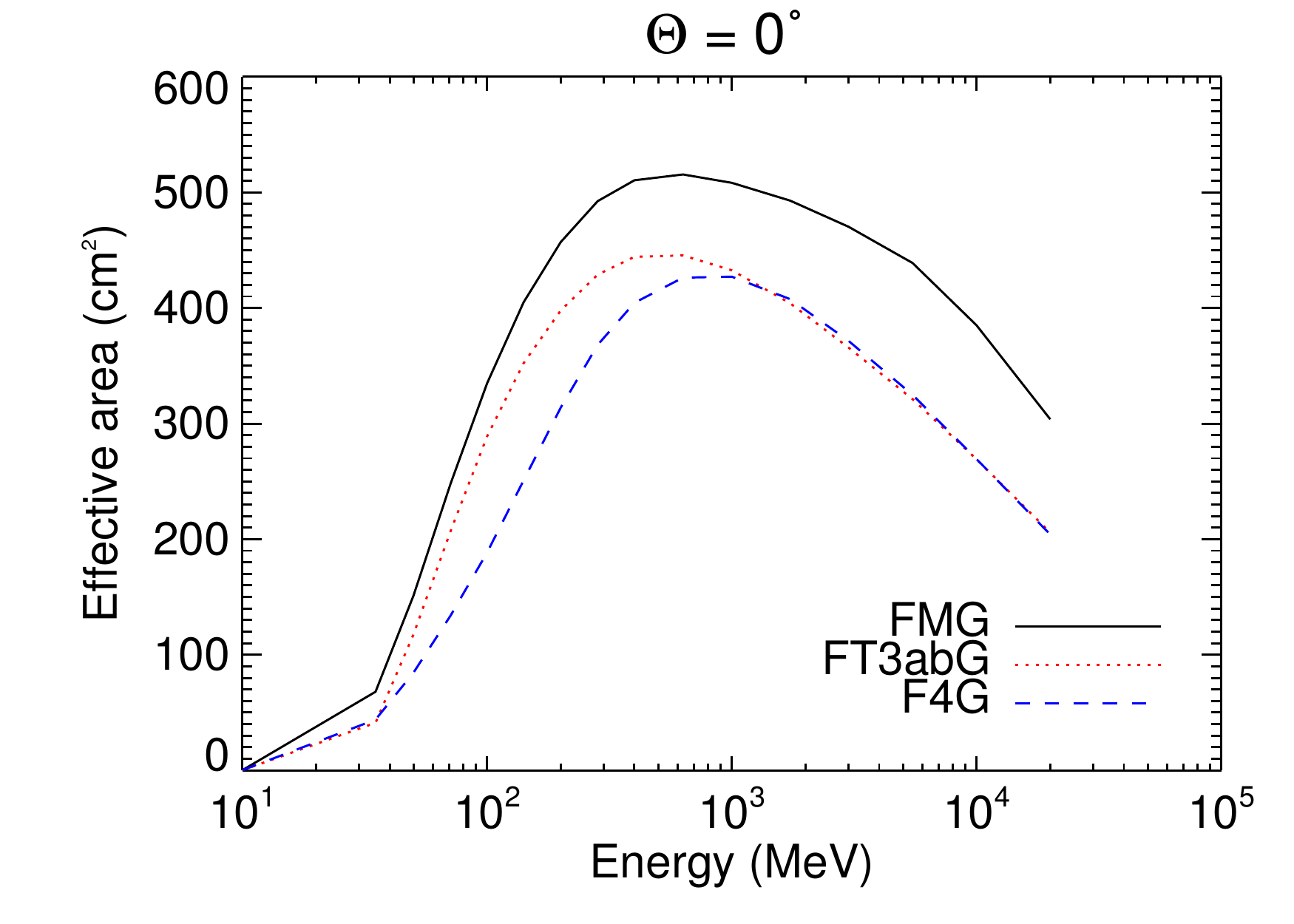}}
    \resizebox{\hsize}{!}{\includegraphics{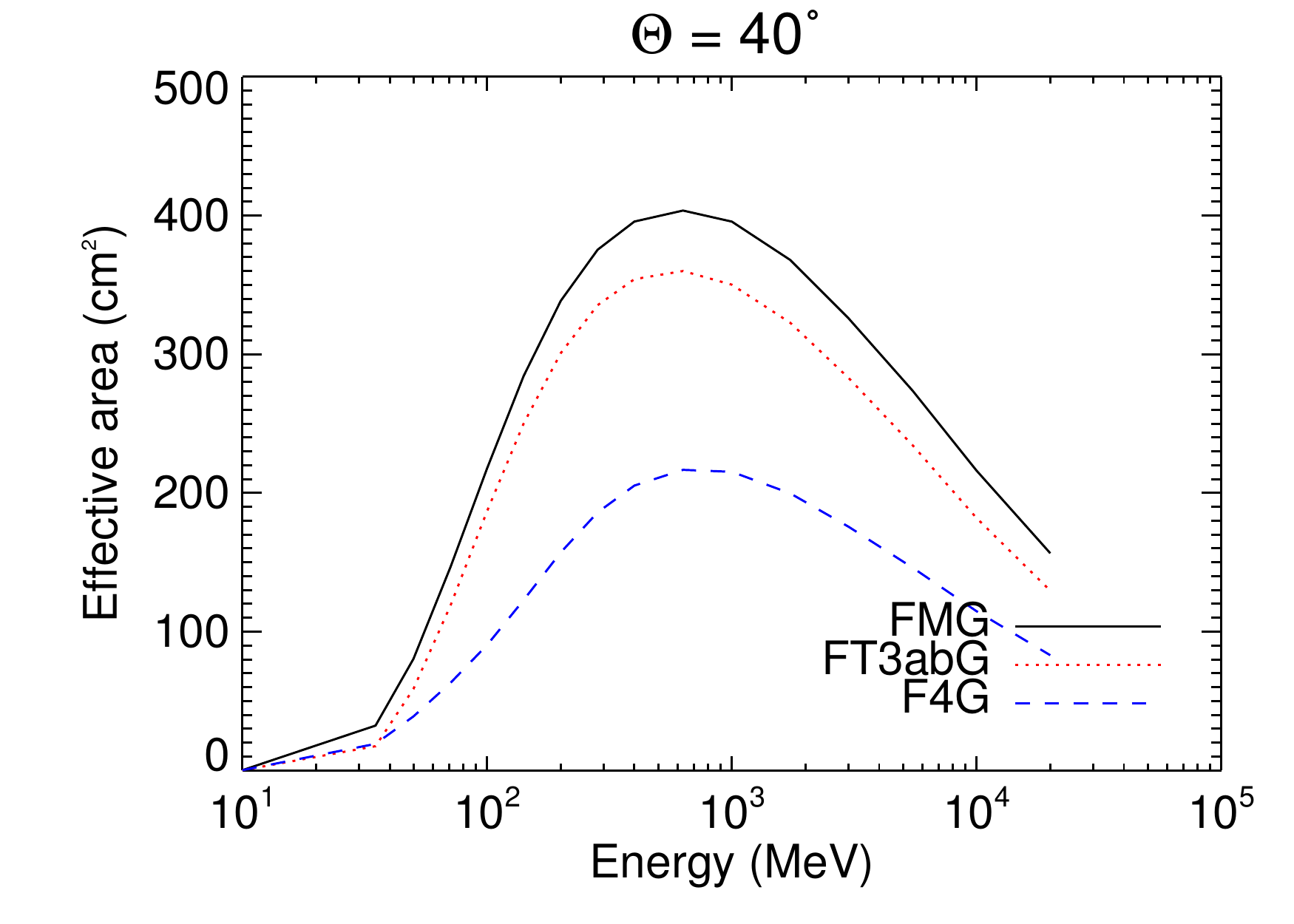}}
    \caption[aeffmono]{\label{aeffmono_lin}
    {AGILE effective areas as a function of energy. Effective area = geometric area $\times$ fraction of surviving events. The top plot is for $\Theta=0^{\circ}$, the bottom plot for $\Theta=40^{\circ}$. AGILE curves are for filters FT3ab and FM3.119, event class G.}}

\end{figure}

\begin{figure}
    \resizebox{\hsize}{!}{\includegraphics{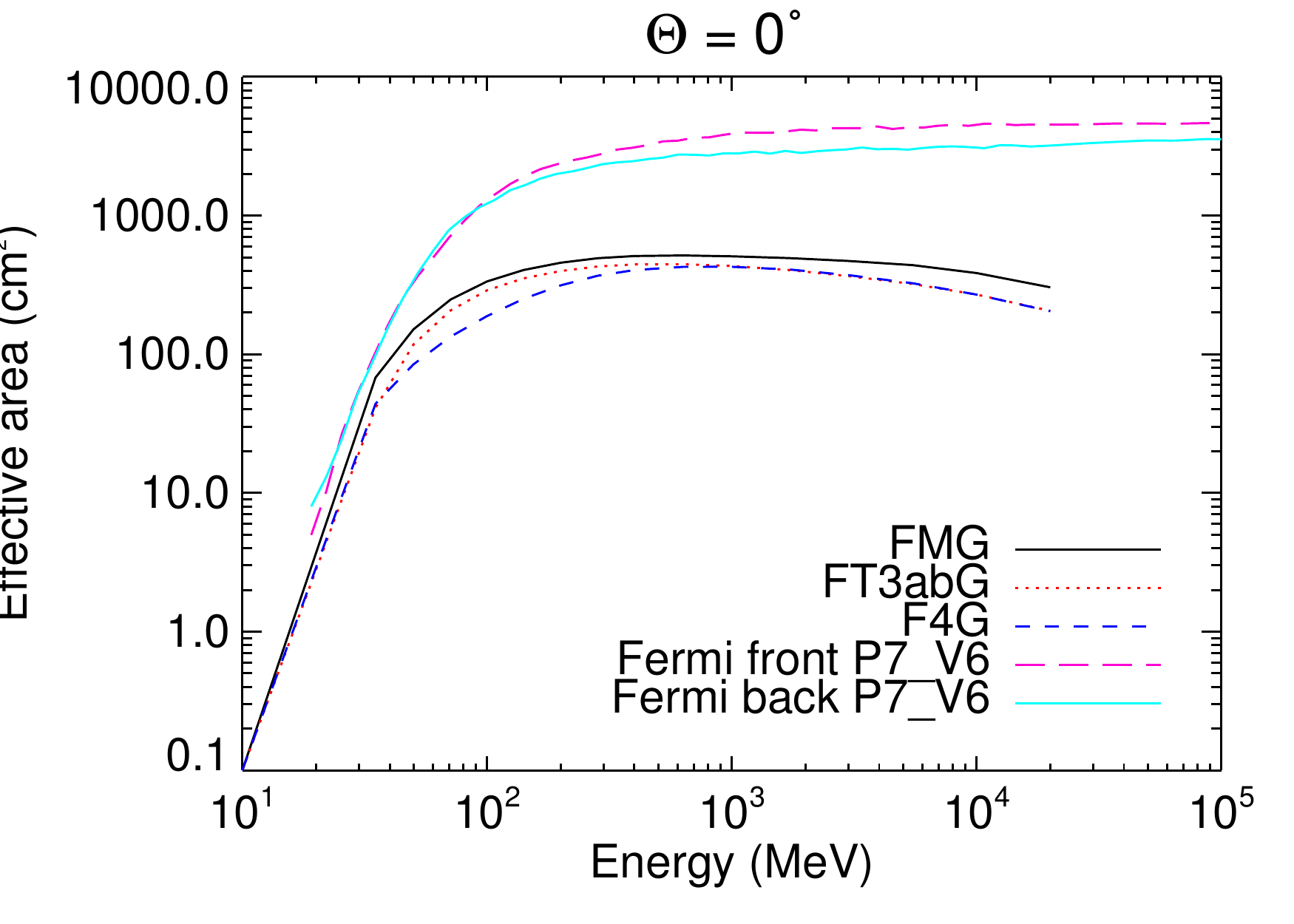}}
    \resizebox{\hsize}{!}{\includegraphics{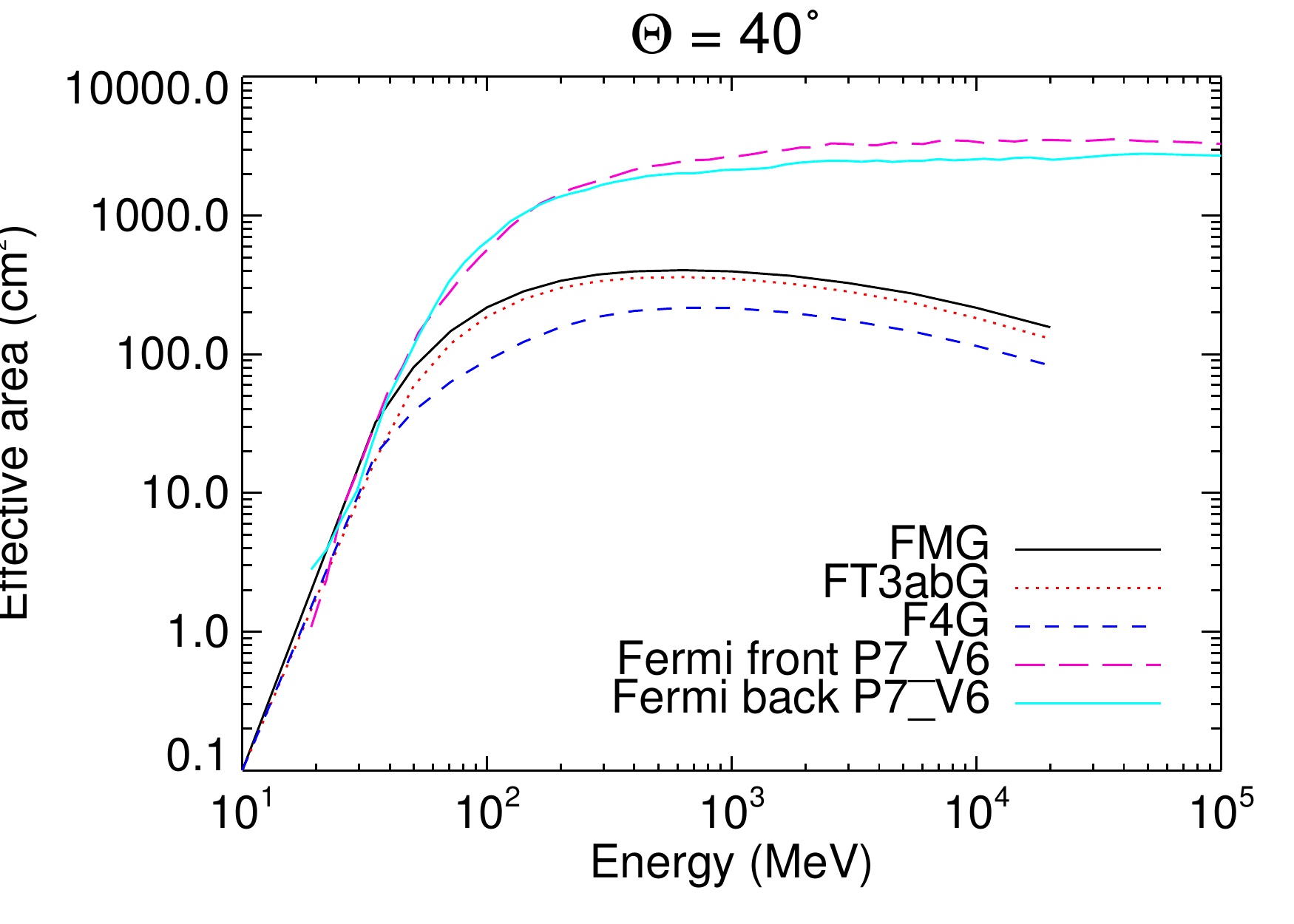}}
    \caption[aeffmono]{\label{aeffmono}
    AGILE and Fermi effective areas as a function of energy. The top plot is for $\Theta=0^{\circ}$, the bottom plot for $\Theta=40^{\circ}$. AGILE curves are for filters FT3ab and FM3.119, event class G. Fermi Pass 7 curves are for version 6, SOURCE event class, front and back events. Fermi IRFs are taken from the Fermi Science Tools, version v9r23p1. The current version is available for public download at \url{http://fermi.gsfc.nasa.gov/ssc/data/analysis/software/}.}

\end{figure}

For the effective area matrix as a function of $\Omega$, the
events for each event class were separated into $N_m=16$ energy bins, whose boundaries are 10, 35, 50, 71, 100, 141, 200, 283, 400, 632, 1000, 1732, 3000, 5477, 10000, 20000, and 50000~MeV. For each energy bin
$i$ containing $\gamma$-rays with energies between $E_i$ and $E_{i+1}$, the number of events classified as event class $V$, is $C(i,\Omega,V)$. The effective area $A_\mathrm{eff}(i,\Omega,V)$ is then defined as
\begin{equation}
A_\mathrm{eff}(i,\Omega,V) =  A_\mathrm{geom}  \frac{C(i,\Omega,V)}{C_\mathrm{tot}}  \frac{E_\mathrm{max} ^ {\alpha+1} -
{E_\mathrm{min} ^ {\alpha+1}}}{E_{i+1} ^ {\alpha+1} - E_{i} ^ {\alpha+1}} 
\end{equation}
where $A_\mathrm{geom}$ is the geometric area of the instrument, $C_\mathrm{tot}$ is the total number of events as defined in Sect. \ref{sec:MC}, $E_\mathrm{max} = E_{N_m+1} = 50$~GeV and $E_\mathrm{min} = E_1 = 10$~MeV.
Some results are shown in Fig. \ref{aeffmono_lin} and compared to the effective area of Fermi-LAT in Fig. \ref{aeffmono} \citepads{2012ApJS..203....4A}.

\subsection{Energy dispersion probability}
\label{sec:edp}

\begin{figure}
   \centering
    \includegraphics[width=8.5cm]{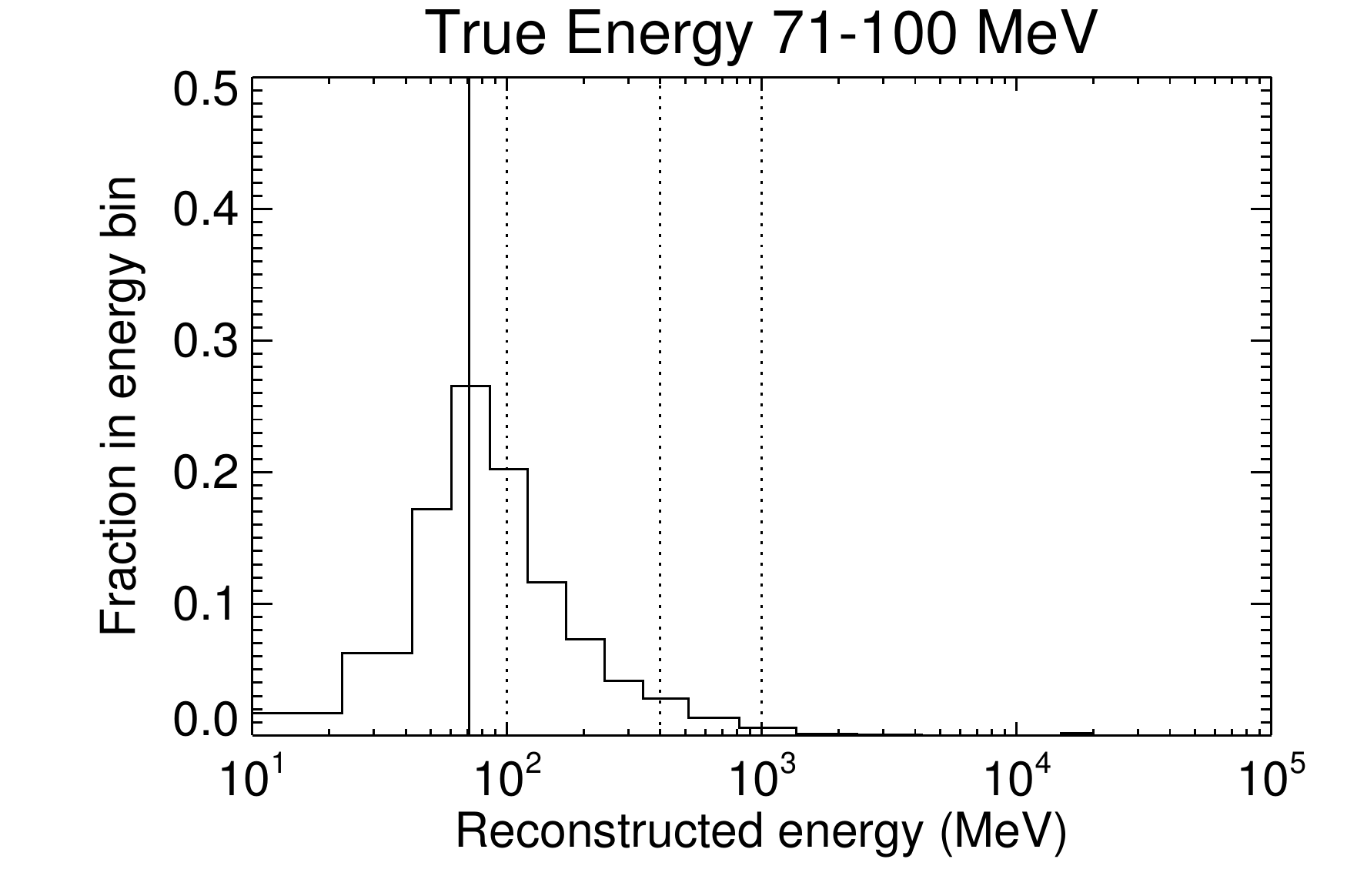}
    \includegraphics[width=8.5cm]{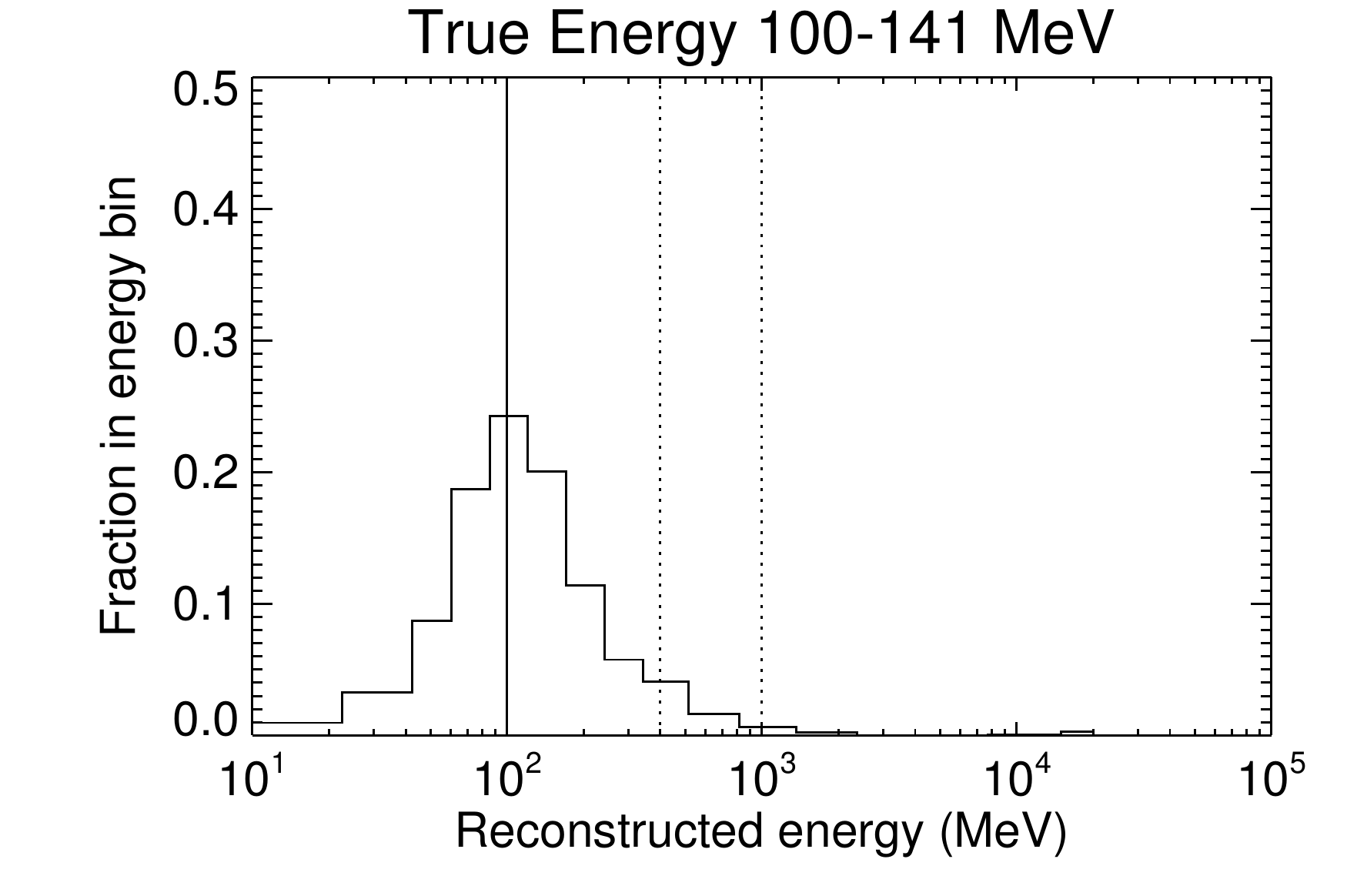}
    \includegraphics[width=8.5cm]{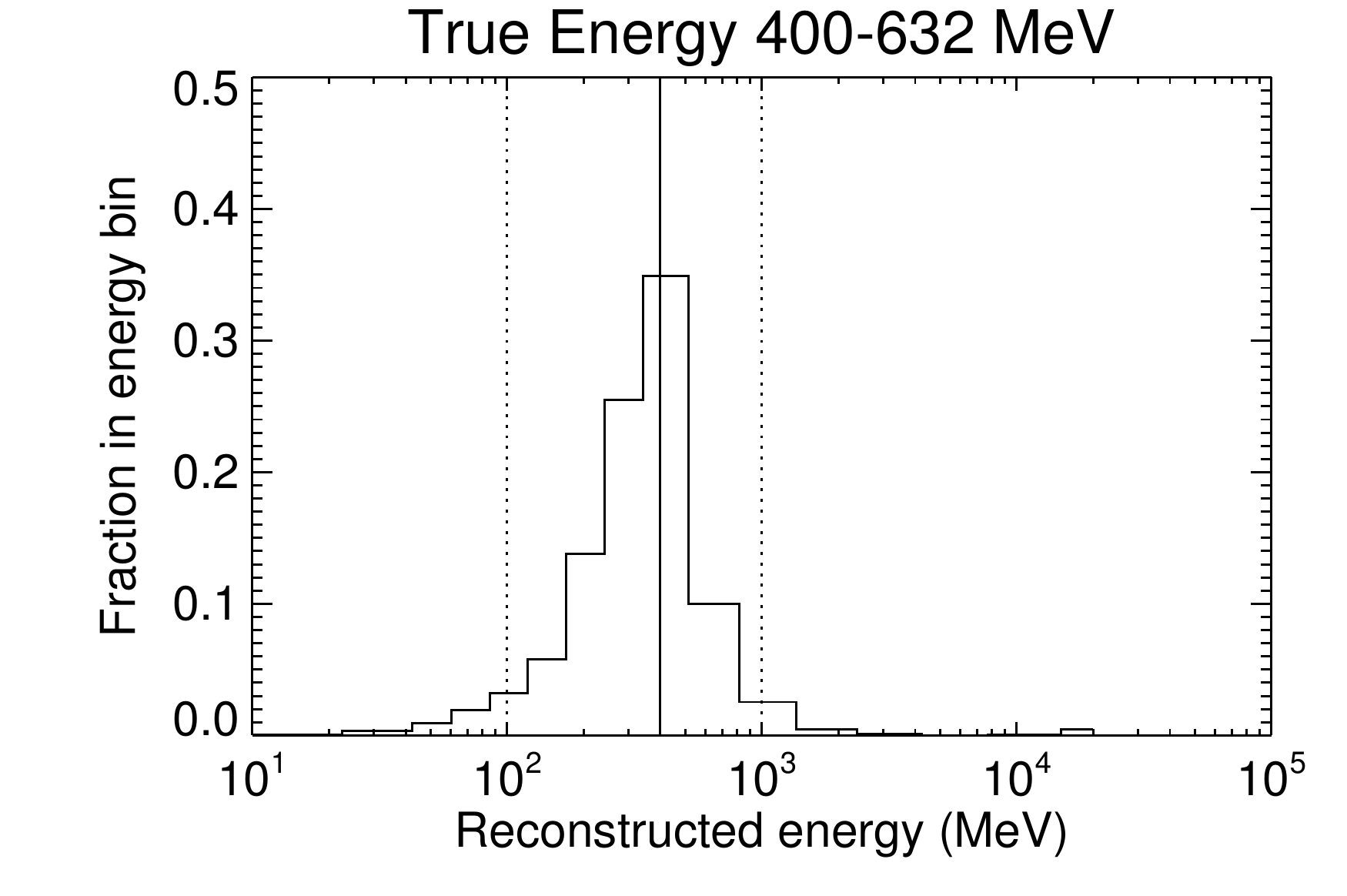}
    \includegraphics[width=8.5cm]{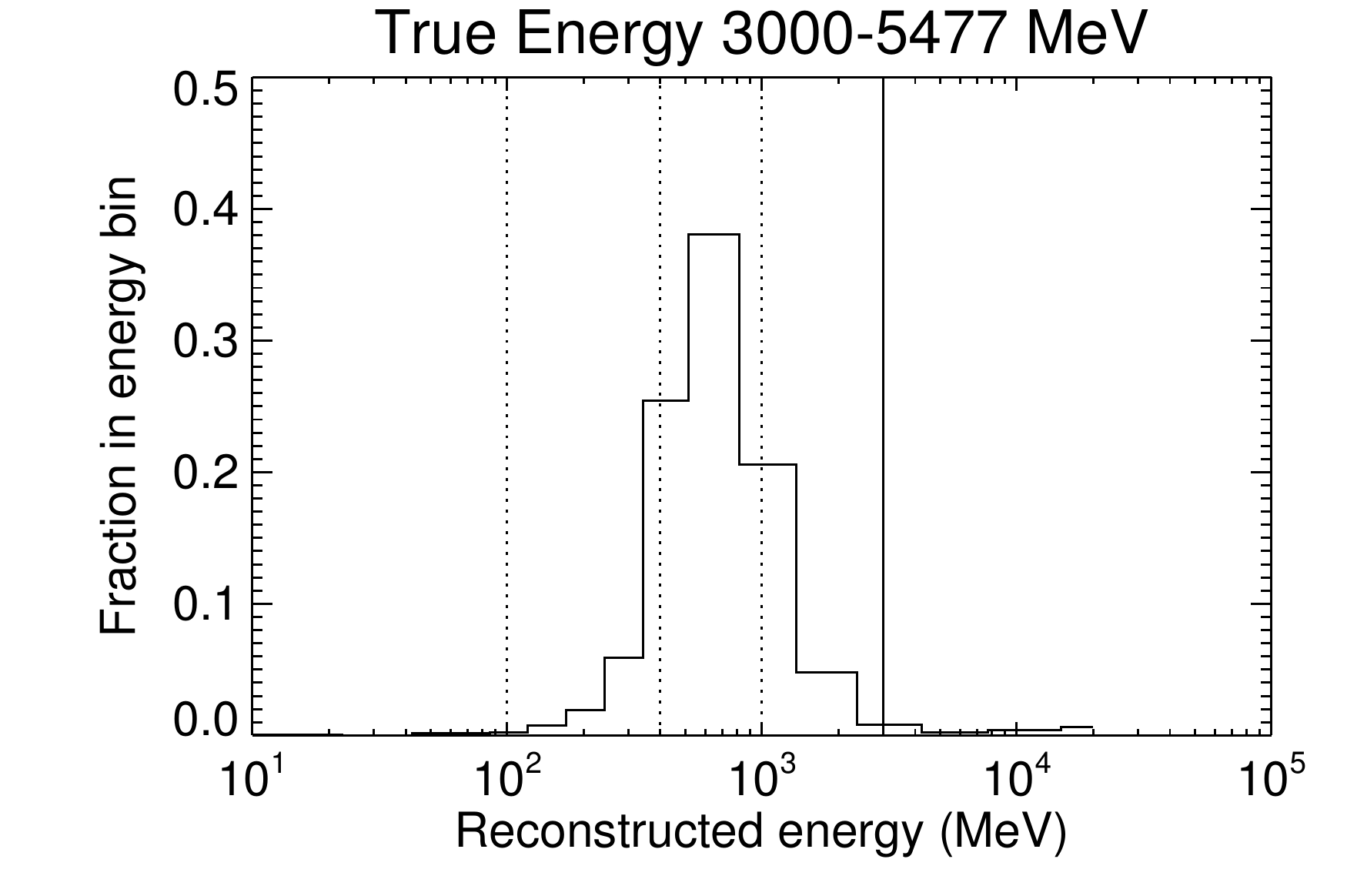}
    \caption{EDPs for filter FMG at various energy bins (71-100, 100-141, 400-632, and 3000-5477 MeV) at $\Theta=30^{\circ}$. Within each bin the true energy follows a power-law distribution (see Sect.~\ref{sec:MC}). The solid vertical line is the lower boundary of the true energy bin, while the dotted lines are fixed at 100, 400, and 1000~MeV.}
\label{EDPfig}
  \end{figure}

\begin{figure}
	\resizebox{\hsize}{!}{\includegraphics{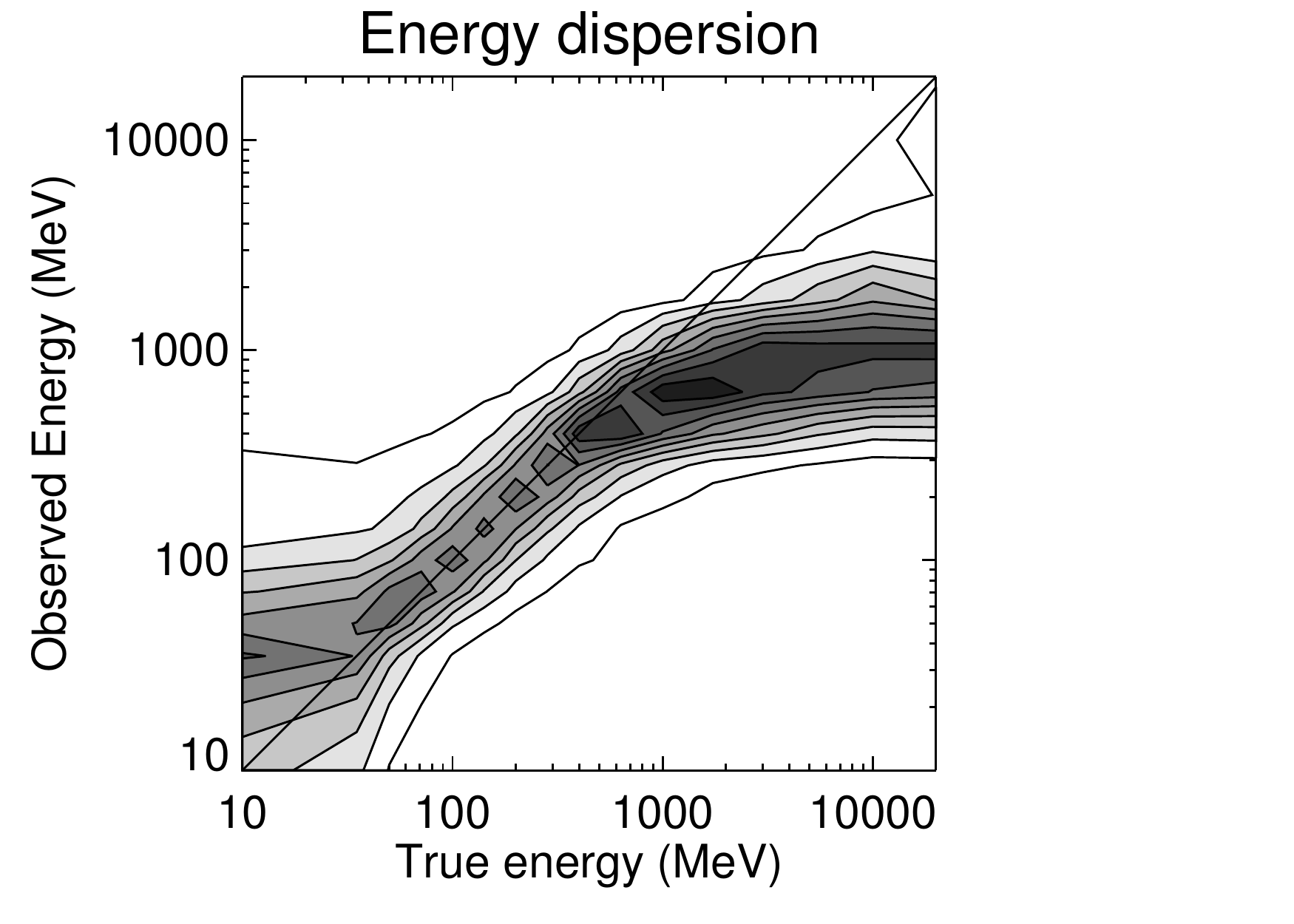}}
	\caption{EDP for filter FMG at $\Theta=30^{\circ}$. Note the deviations from linearity below 100~MeV and above 400~MeV.}
\label{EDPgraph}
\end{figure}

The AGILE energy dispersion matrices use the same energy bins for the true
and reconstructed energies. For each event class and set of instrument coordinates,
the EDP is the fraction of events within a given true
energy bin whose reconstructed energy lies within a given reconstructed
energy bin. The EDPs for the G event class of the FM3.119 filter
(hereafter referred to as FMG) for selected energy bins at $\Theta=30^{\circ}$ are shown in
Fig. \ref{EDPfig}. Note that a substantial fraction of $\gamma$-rays with true energy below 100~MeV
have reconstructed energies above 100~MeV, implying that a substantial fraction of
events with reconstructed energies above 100~MeV will have true energies
below 100~MeV for most astrophysical $\gamma$-ray sources, which tend to
have spectral indices $\alpha \approx -2$. Any $\gamma$-ray source which emits
primarily below 100~MeV will also be detected in the nominal $E > 100$~MeV band.
Meanwhile, a majority of $\gamma$-rays with true energy above 1~GeV
have reconstructed energies below 1~GeV. Any $\gamma$-ray source which emits
primarily above 1~GeV will have most of its flux reconstructed in the 400~MeV $< E < 1000$~MeV band. Both of these effects are due to the limitations of multiple scattering as the primary method of energy reconstruction; at lower energies, a certain fraction of events will nevertheless be scattered at small angles (where the peak of the angular distribution lies; see the description in the next section), while at high energies 
the pitch of the silicon microstrips, 121~\mbox{$\mu$m}, is too coarse to measure the scattering angle and the Mini-Calorimeter reaches its saturation point.
The relationship between true and observed energy is shown in Figs. \ref{EDPfig} and \ref{EDPgraph}.
The AGILE-GRID analysis software takes these factors into account, but discrepancies
may arise if the spectral index is fixed to the wrong value or if the spectrum
diverges significantly from a power law.

\subsection{Point spread function}

\begin{figure}
    \resizebox{\hsize}{!}{\includegraphics{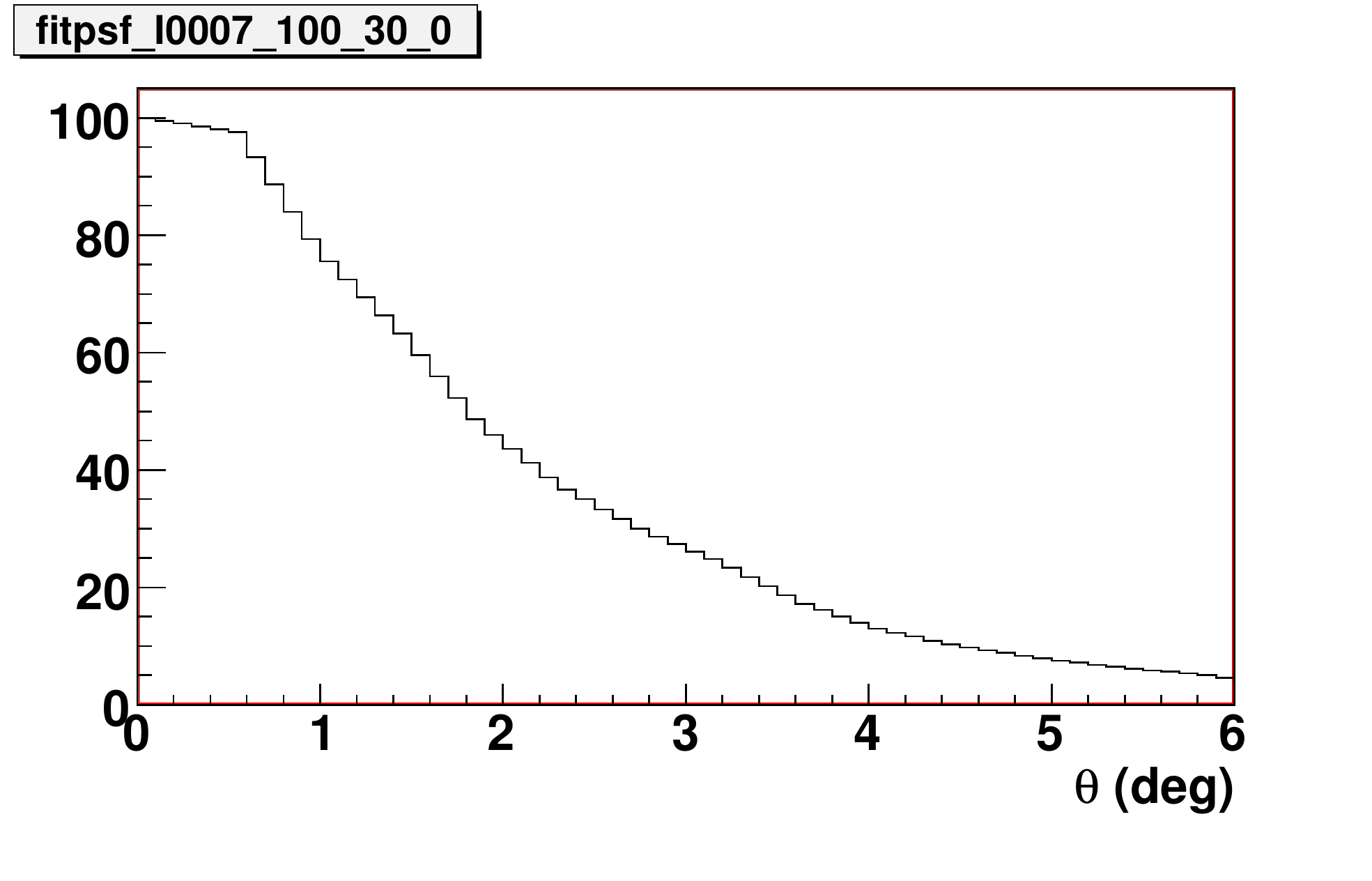}}
    \resizebox{\hsize}{!}{\includegraphics{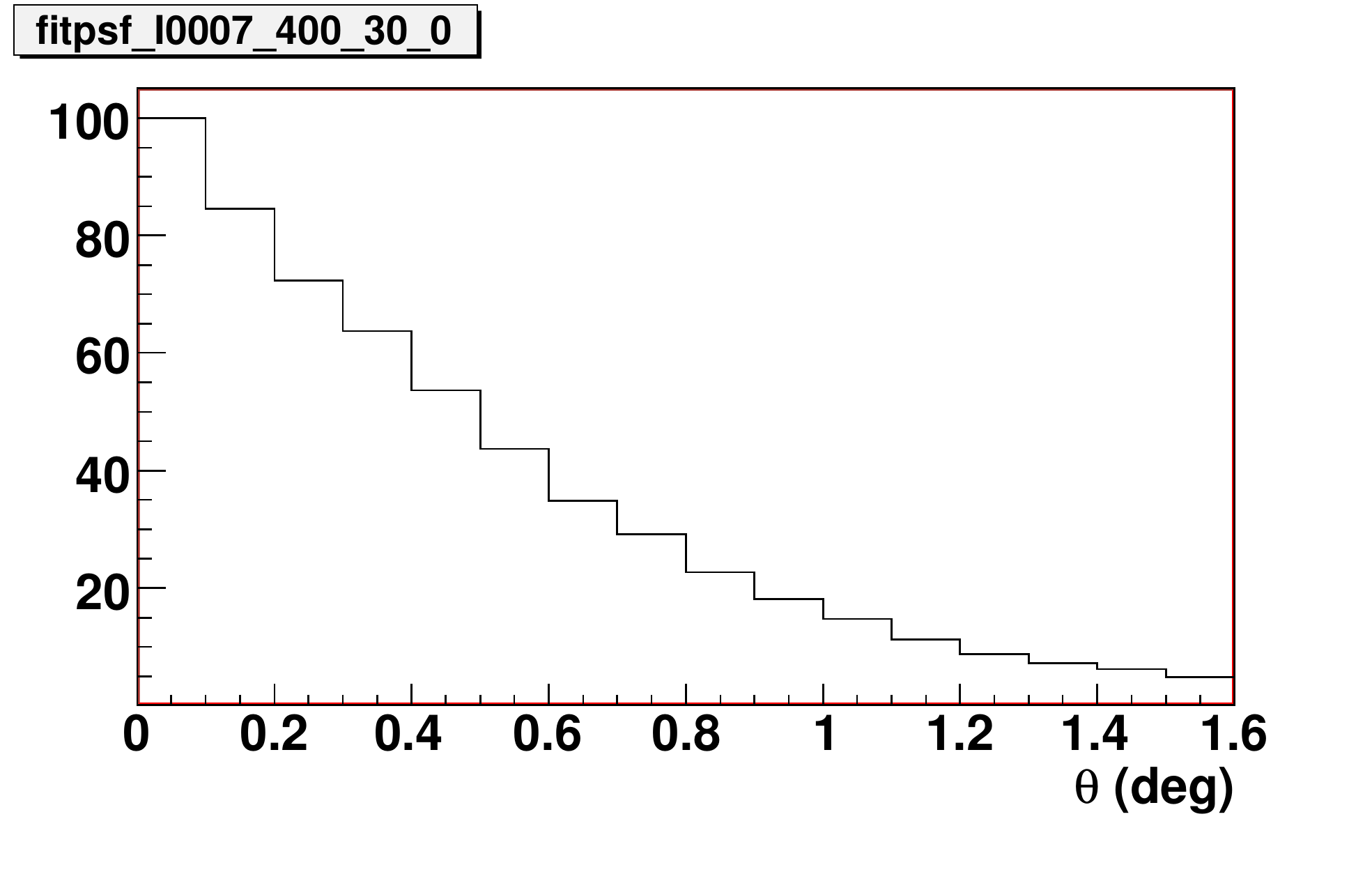}}
    \resizebox{\hsize}{!}{\includegraphics{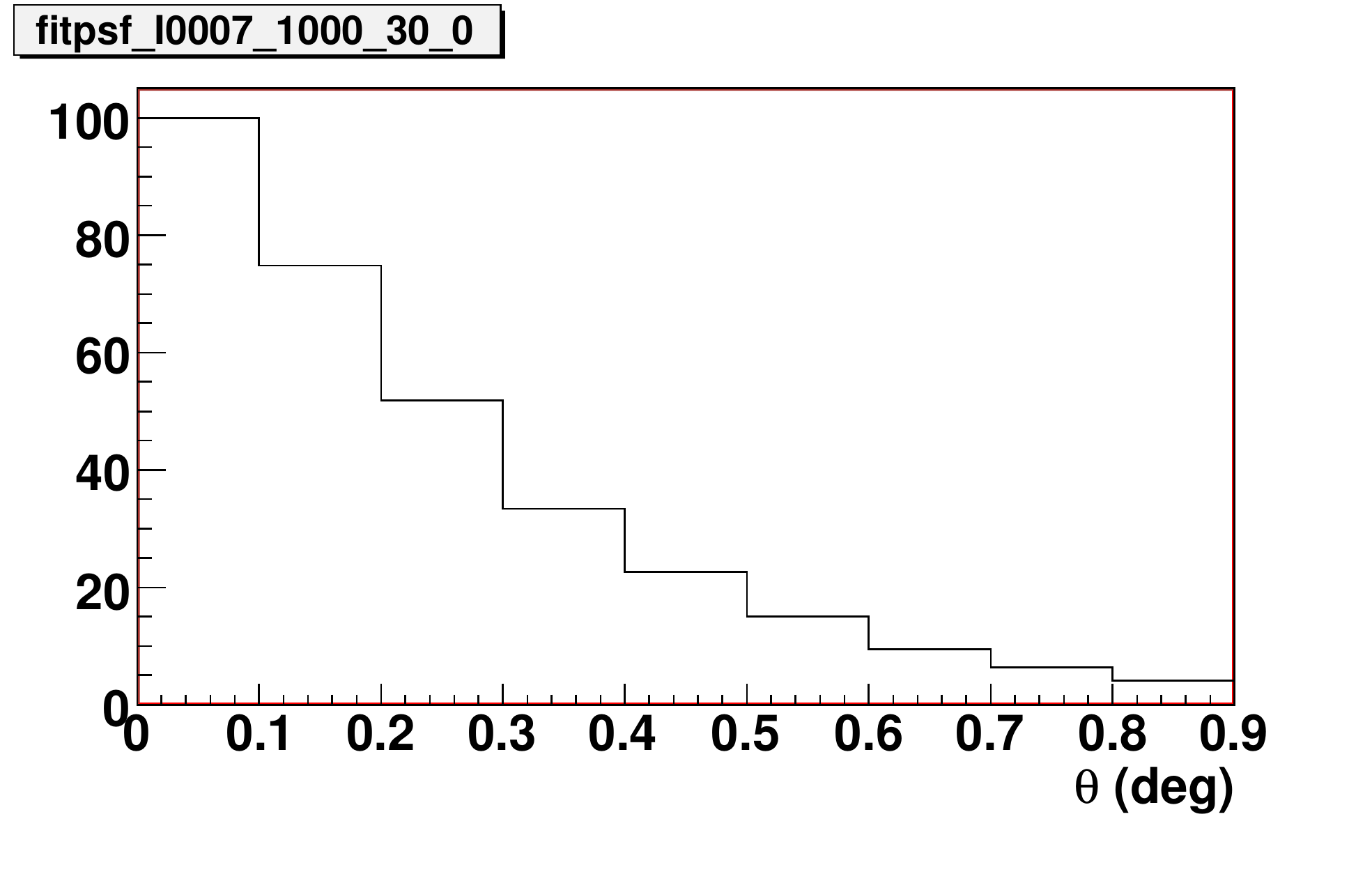}}
    \caption{Monoenergetic PSFs at $\Theta=30^{\circ}$ with filter FMG at 100, 400, and 1000~MeV. The I0007 PSF matrices were created by directly binning the Monte Carlo data, dividing the raw histogram by $\sin \theta$, and normalizing. The I0010 PSF matrices are identical to those of I0007.}
    \label{PSFfig}
\end{figure}

\begin{figure}
    \resizebox{\hsize}{!}{\includegraphics{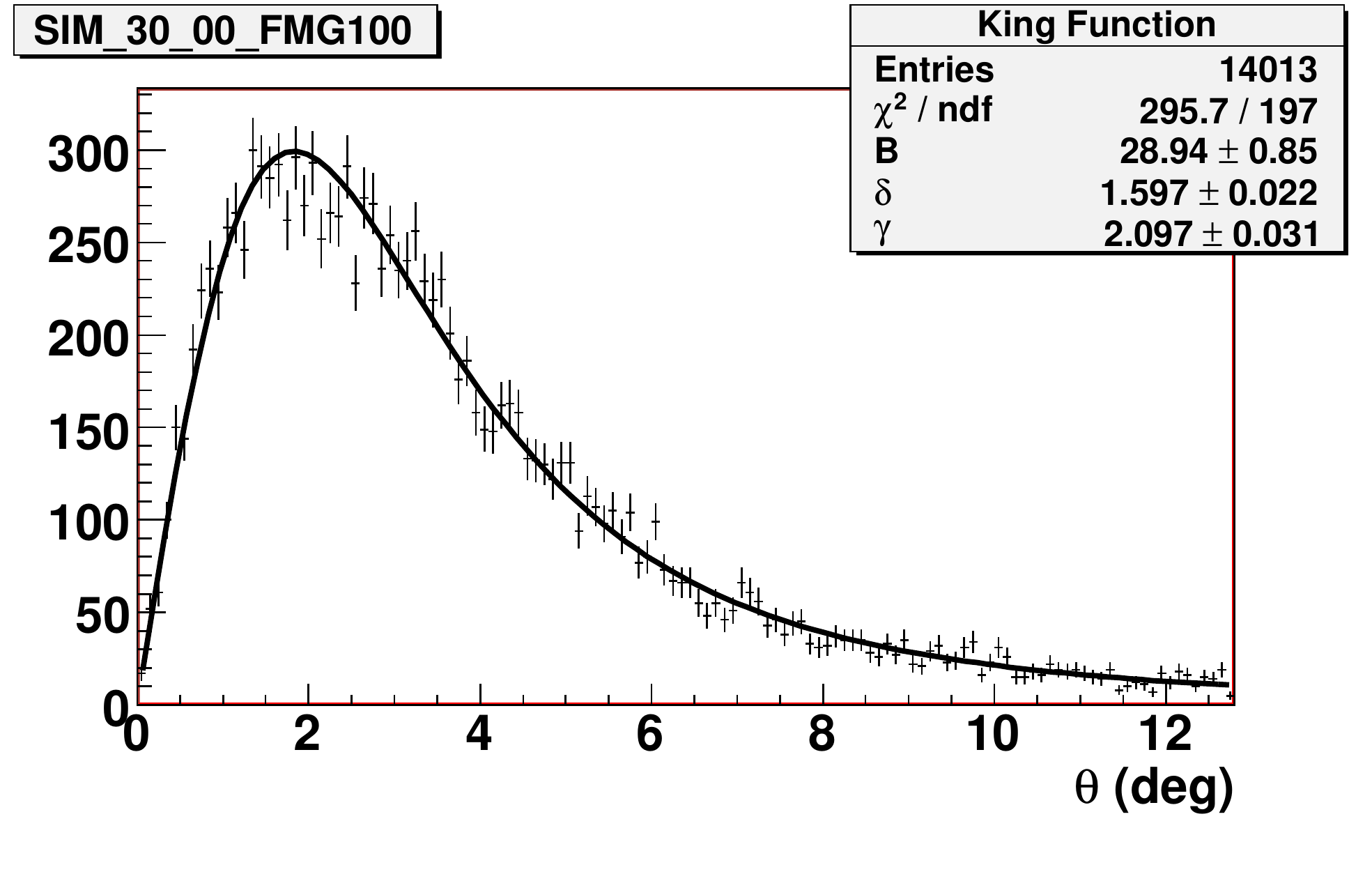}}
    \resizebox{\hsize}{!}{\includegraphics{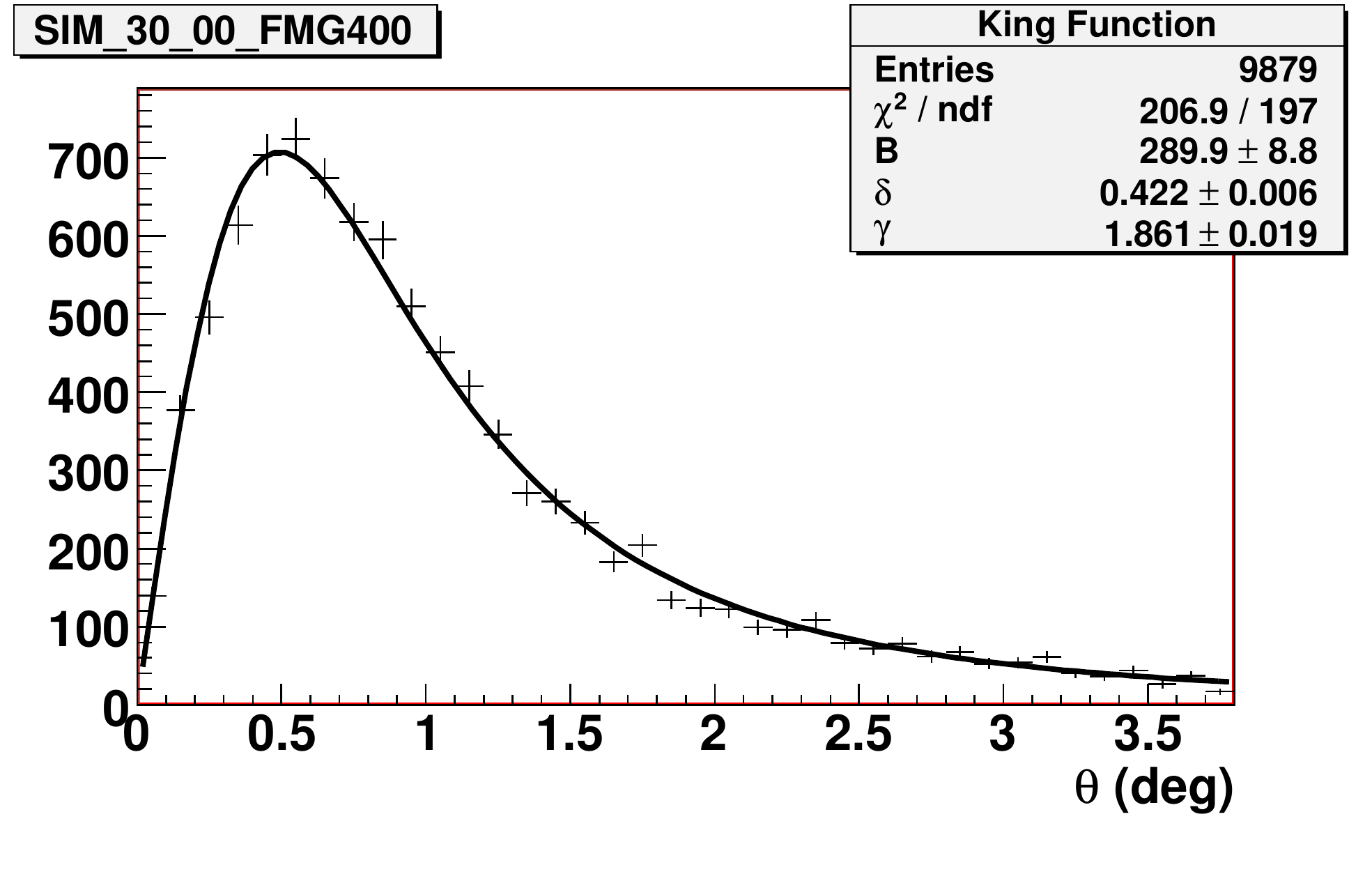}}
    \resizebox{\hsize}{!}{\includegraphics{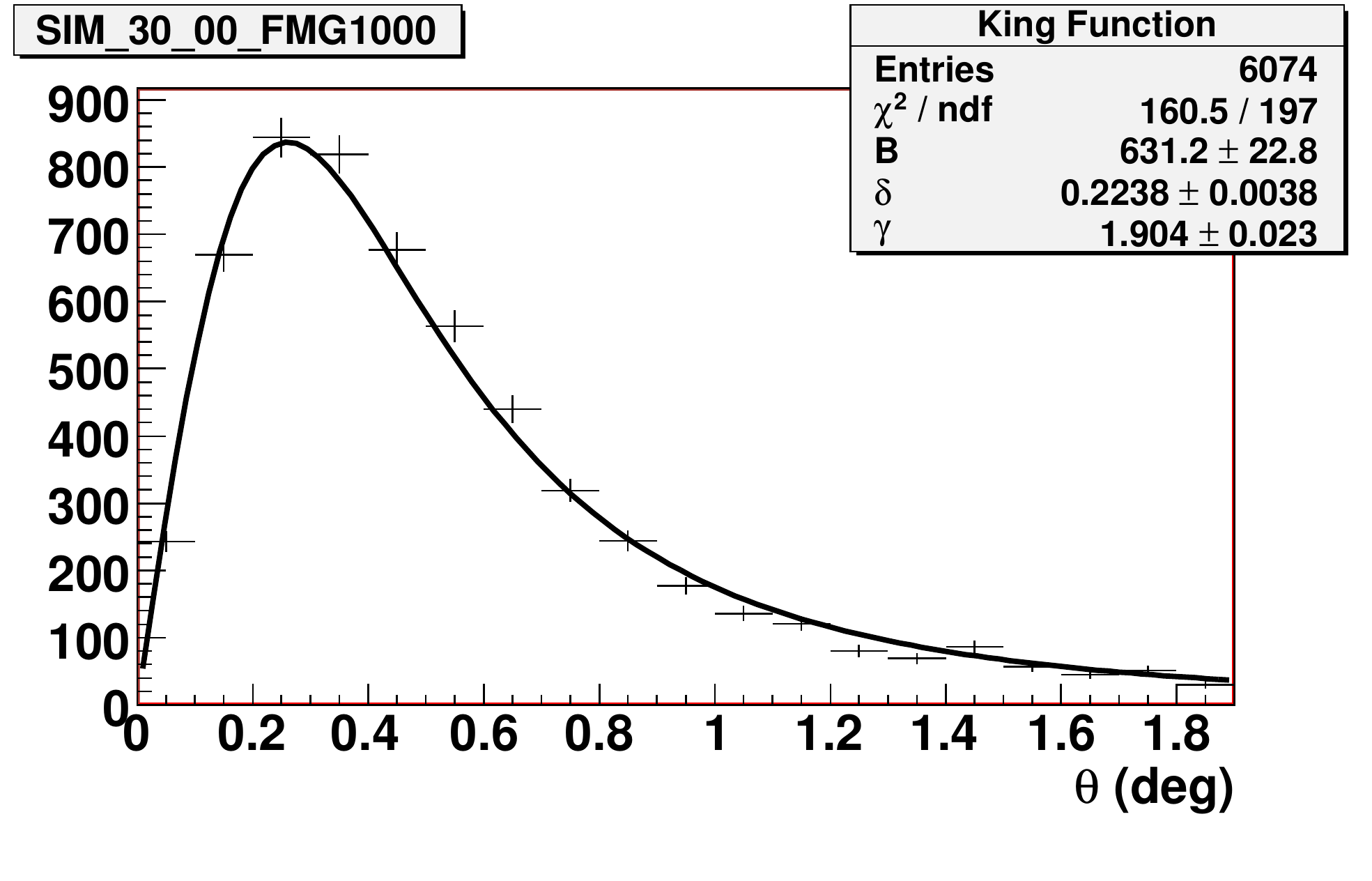}}
    \caption{Monoenergetic PSFs at $\Theta=30^{\circ}$ with filter FMG at 100, 400, and 1000~MeV. The I0023 PSF matrices were created by first fitting 
Eq. (\ref{kingfunction}) to the Monte Carlo data, then binning the values of the King function and normalizing.}
    \label{PSFfig_I0023}
\end{figure}

A series of $\gamma$-rays from the same direction in instrument coordinates will have a distribution of reconstructed directions, an effect known as Point Spread Dispersion (PSD). The PSF, which also depends on the $\gamma$-ray energy and event class, is defined as the probability distribution of the angular distance $\theta$ between the reconstructed and the true direction. The PSF is estimated from Monte Carlo simulations. Some examples of PSFs are shown in Figs. \ref{PSFfig} and \ref{PSFfig_I0023}.

\section{Fluxes, spectra and PSFs of real sources}

The flux and spectrum of any physical point source can be decomposed into a series of monoenergetic point sources
whose fluxes are equal to the differential flux of the point source at each energy. Each monoenergetic point source has a well-defined
$A_\mathrm{eff}$, EDP, and PSF. These quantities
are used to calculate the composite effective area and PSF of the physical source depending on its spectrum and
coordinates in the instrument frame.

Several versions of the IRFs have been used for the AGILE-GRID analysis. 
Version I0007, used internally since the beginning of 2009 and released publicly on May 22, 2009 
in software release 3.0 by the AGILE Data Center\footnote{\url{http://agile.asdc.asi.it/}}, 
part of the ASI Science Data Center (ASDC),
used histograms directly binned from Monte Carlo data for the PSFs, 
without fitting to any analytic function. 
Version I0010, used internally from August 2009 until the end of 2010 and never released in public software packages,
used the same directly binned PSFs, but introduced correction factors into the 
effective area matrices in a first attempt to account for energy dispersion when 
calculating the effective area for real sources. 
Finally, the latest version (I0023), used internally by the AGILE team since the end of 2010,
and publicly included in ASDC software release 5.0 on November 25, 2011, 
fills the PSFs with an analytic King function fit to the Monte Carlo data, 
while removing the effective area correction factors introduced into I0010. 
A new exposure generation procedure which accounts for energy dispersion 
will be included in an upcoming software release. 
These characteristics are summarized in Table \ref{table:IRFs}.

\begin{table}
  \caption[]{Versions of the AGILE-GRID IRFs.}
  \label{table:IRFs}
  \centering
  \begin{tabular}{c | c c c}
    \hline \hline
    IRF version & I0007 & I0010 & I0023 \\
    \hline \hline
    Internal validation & 2009-02-18 & 2009-08-03 & 2010-11-22\\
    begins & & & \\
    \hline
    Public software & 2009-05-22 & --- & 2011-11-25\\
    release & & & \\
    \hline \hline
    PSF binned directly & X & X &  \\
     from MC histograms & & & \\
    \hline
    PSF binned  from &  &  &  X \\
    fit to King function & & & \\
    \hline \hline
    Correction factors  & No & Yes & No \\
    applied to eff. area & & & \\
    \hline
  \end{tabular}
\end{table}

\subsection{Effective area}
\label{sec:Aeff2}

Suppose that a $\gamma$-ray source has a power-law spectrum
$dN/dE = N E^{\alpha}$. Then the flux in the energy bin $i$ is 
\begin{equation}
	\label{trueflux}
	F_i=N\int_{E_i}^{E_{i+1}}E^{\alpha}dE=\frac{N}{\alpha+1}(E_{i+1}^{\alpha+1}-E_i^{\alpha+1})
	\end{equation}
	and the total flux between energies $E_a$ and $E_b$ is 
	\begin{equation}\label{truefluxab}
	 F_{ab} = \sum_{i=a}^{b-1} F_i=\frac{N}{\alpha+1}(E_b^{\alpha+1}-E_a^{\alpha+1}).
\end{equation}
If an instrument with $A_\mathrm{eff}(i,\Omega,V)$ for $\gamma$-rays whose true energy lies within energy bin $i$ is exposed to the source for time $t$, the number of counts in each energy bin is
\begin{equation}
C_i(\Omega,V)=F_i A_\mathrm{eff}(i,\Omega,V) t.
\end{equation}
If $EDP(i,j,\Omega,V)$ is the fraction of $\gamma$-rays whose true energy lies in energy bin $i$ which have reconstructed energy within energy bin $j$, the number of counts from the full source spectrum whose reconstructed energy lies in energy bin $j$ is
\begin{equation}
C'_j(\Omega,V)=\sum_{i=0}^{N_m}C_i EDP(i,j,\Omega,V)
\end{equation}
with $N_m$ defined as in Sect. \ref{sec:Aeff}.
Therefore the total number of counts whose observed energies lie between $E_a$ and $E_b$ is
\begin{multline}\label{obscounts}
C'_{ab}(\Omega,V)=\sum_{j=a}^{b-1}C'_j(\Omega,V)\\
=\sum_{j=a}^{b-1}\sum_{i=0}^{N_m}\frac{N}{\alpha+1} A_\mathrm{eff}(i,\Omega,V) t EDP(i,j,\Omega,V)(E_{i+1}^{\alpha+1}-E_i^{\alpha+1})
\end{multline}
where both the effective areas and EDPs for individual energy bins and the observed effective areas are functions of the $\gamma$-ray direction $\Omega$ in instrument coordinates and
event type $V$.

The effective area with respect to an interval of observed energies $E_a$ and $E_b$ is defined as the number of counts whose observed energies lie between $E_a$ and $E_b$ (Eq. \ref{obscounts}) divided by the true flux between $E_a$ and $E_b$ (Eq. \ref{truefluxab}) divided by the time of observation $t$ as follows:
\begin{multline}\label{trueaeffeq}
A'_{ab}(\Omega,V)=\frac{C'_{ab}(\Omega,V)}{F_{ab}t}\\
=\frac{\sum_{i=0}^{N_m} A_\mathrm{eff}(i,\Omega,V) (E_{i+1}^{\alpha+1}-E_i^{\alpha+1})\sum_{j=a}^{b-1}EDP(i,j,\Omega,V)}{E_b^{\alpha+1}-E_a^{\alpha+1}}.
\end{multline}

Note that $A'_{ab}$ can be expressed as a weighted sum of $A_\mathrm{eff}(i,\Omega,V)$ as follows:
\begin{equation}
A'_{ab}(\Omega,V)=\sum_{i=0}^{N_m}A_\mathrm{eff}(i,\Omega,V)\frac{E_{i+1}^{\alpha+1}-E_i^{\alpha+1}}{E_b^{\alpha+1}-E_a^{\alpha+1}}w_{ab}(i,\Omega,V)
\end{equation}
where
\begin{equation}
w_{ab}(i,\Omega,V)=\sum_{j=a}^{b-1}EDP(i,j,\Omega,V)
\end{equation}

As of this writing, a simpler formula for the energy weight, not taking
into account the EDPs,
has been used, where the scaling factors $w_{ab}(i,\Omega,V)$ were
set equal to 1 for the first version of
the IRFs (I0007),
and determined post-hoc as a function of instrument coordinates $\Omega$ according to the procedure in
Sect. \ref{Flux and spectra}. These post-hoc scaling factors were incorporated directly into the effective area matrices in version I0010 of the IRFs.

However, we have found too limited the range of spectral indices for which
this simplified formula is applicable, and are implementing the correct formula in the soon-to-be-released BUILD 22 of the software.

\subsection{Point spread function}
\label{sec:psf}

The PSF for a physical source observed in an interval of reconstructed energies is the weighted average of the PSFs in individual
energy bins, where the weight of each energy bin is proportional to the product of the effective area, the flux in the energy bin (determined by the source spectrum), and the fraction of $\gamma$-rays from the energy bin whose reconstructed energy lies within the observed reconstructed energy interval (determined by the EDP).
If the source has power-law index $\alpha$ between energies $E_a$ and $E_b$, the PSF is
\begin{equation}
{PSF}'_{ab}(\Omega,V) = \frac{\sum_{i=0}^{N_m} q_{ab}(i,\Omega,V) {PSF}(i,\Omega,V)}{\sum_{i=0}^{N_m} q_{ab}(i,\Omega,V)}
\label{psfeq}
\end{equation}
where
\[
 q_{ab}(i,\Omega,V) = A_\mathrm{eff}(i,\Omega,V)(E_{i+1}^{\alpha+1} - E_i^{\alpha+1})   \sum_{j=a}^{b}{EDP}(i,j,\Omega,V).
\]

Earlier versions (I0007/I0010) of the PSF matrices used histograms
taken directly from the Monte Carlo simulations. The updated PSF matrices (I0023)
contain values derived from a fit to the Monte Carlo
data using a modified King function \citepads{1962AJ.....67..471K} used to characterize high-energy PSFs (\citeads{2004SPIE.5488..103K}; \citeads{2011A&A...534A..34R}) $f(\theta)$, which has three parameters, $B$, the (arbitrary) normalization, $\delta$, the characteristic width, and $\gamma$, which is related to the relative strength of the core vs. the tail, as follows:
\begin{equation}
\label{kingfunction}
 f(\theta)\sin \theta\,\mathrm{d}\theta =
B(1-1/\gamma)\left(1+\frac{(\theta/\delta)^2}{2\gamma}\right)^{-\gamma}\sin \theta\,\mathrm{d}\theta
\end{equation}

The PSF matrices are then filled with the values derived
from the King function with a bin size of $0.1^{\circ}$.

We compare the 68\% $\gamma$-ray Containment Radii (CRs) of the PSFs in single, true energy bins (Table \ref{psftab1}) with those of the composite PSFs in broad, reconstructed energy intervals (Table \ref{psftab2}).
Note that the CR for the reconstructed $E > 1$~GeV interval is broader than that of the true $E = 1$~GeV bin. This is because, as we showed in Sect. \ref{sec:edp}, the reconstructed $E > 1$~GeV interval is dominated by $\gamma$-rays whose true energy is actually below 1~GeV. 

\begin{table}
         \caption{68\% $\gamma$-ray Containment Radii (CRs) of monoenergetic PSFs.
        \label{psftab1}}
        \centering
	\begin{tabular}{cc}
        \hline \hline
	True energy (MeV) & 68\% CR \\
	\hline
	100 & $4.3^{\circ}$  \\
	400 & $1.4^{\circ}$ \\
	1000 & $0.7^{\circ}$\\
        \hline
	\end{tabular}
	 \tablefoot{Monoenergetic PSFs for three true energies at $\Theta= 30^{\circ}$ from Monte Carlo data.}
\end{table}

\begin{table}
        \caption{68\% $\gamma$-ray Containment Radii (CRs) of composite PSFs.
        \label{psftab2}}
	\begin{center}
	\begin{tabular}{cc}
        \hline \hline
	Energy Range & 68\% CR \\
	\hline
	100~MeV - 50~GeV & $2.1^{\circ}$  \\
	400~MeV - 50~GeV & $1.1^{\circ}$ \\
	1000~MeV - 50~GeV & $0.8^{\circ}$ \\
        \hline
	\end{tabular}
	 \tablefoot{Composite PSFs for three reconstructed energy intervals at $\Theta= 30^{\circ}$ and spectral index $\alpha = -1.66$ from Monte Carlo data.}
	\end{center}
\end{table}

\section{Comparison to in-flight data}

We generated long-term integrations of AGILE-GRID in-flight data in both
pointing (2007/07/09 - 2009/10/15) and spinning (2009/11/04 - 2010/10/31) modes
of the \object{Vela} and anti-center regions, generating counts and exposure maps with a
bin size of $0.3^{\circ}$.
The AGILE maximum likelihood analysis \citepads{2012A&A...540A..79B} was performed taking into account the Galactic diffuse
emission and the isotropic background, and the following bright point sources:
the \object{Vela} point source, which comprises both the pulsar and the pulsar wind nebula (PWN), and the \object{Crab} and \object{Geminga}
point sources and IC443 in the anti-center region, where the \object{Crab} point source also comprises both the pulsar and the PWN, all with fixed source locations
and fixed, power-law spectra.
Model counts were compared to data to validate the PSF, while
spectra and fluxes were compared to those published in the 
Third EGRET catalog \citepads[][hereafter 3EG]{1999ApJS..123...79H}
in order to determine the post-hoc scaling factors introduced in \ref{sec:Aeff2} that were incorporated into the I0010 effective area matrices and calculated according to the procedure described in the following subsection.

\subsection{Fluxes and spectra: correction factors}\label{Flux and spectra}

  \begin{figure}
    \resizebox{\hsize}{!}{\includegraphics{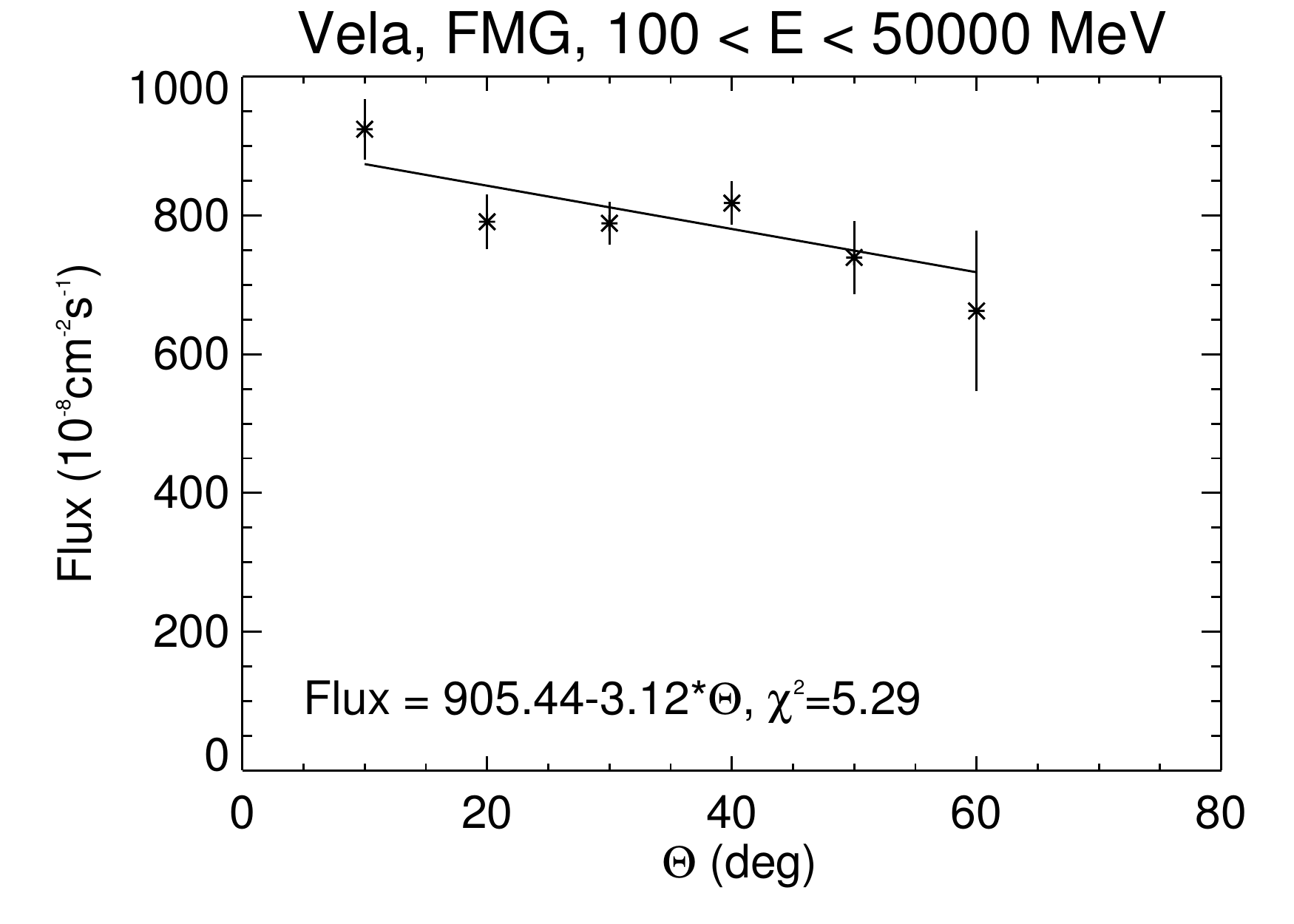}}
    \caption{Observed flux of the source at the position of the \object{Vela pulsar} for E $> 100$ MeV as a function of $\Theta$ with the I0007 IRFs. The linear fit to these fluxes were used to calculate the effective area correction factors in the I0010 IRFs.
    }
    \label{correctionfit}
  \end{figure}

To create the I0010 version of the effective area matrices, we compared the fluxes for E $> 100$ MeV obtained using the I0007 effective areas with the AGILE likelihood analysis of the \object{Vela pulsar} at different off-axis angles with those
expected from the fluxes and spectra reported in the Fermi Large Area Telescope First Source Catalog \citepads[][hereafter 1FGL]{2010ApJS..188..405A}. A linear fit was performed on the fluxes produced by the analysis (Fig. \ref{correctionfit}). The correction factors
were set equal to the inverse of the ratio between the
fluxes implied by the fit parameters and the 1FGL fluxes for $\Theta<60^{\circ}$ and set equal to the value at $60^{\circ}$ for 
$\Theta\ge 60^{\circ}$.
These were applied to the original effective areas to produce new effective areas to be used in
AGILE analysis.

However, when attempting to reproduce this procedure for the updated IRFs, we discovered that
the fluxes and spectra of the softer spectrum of the \object{Crab} were overestimated. In fact, the likelihood analysis of the \object{Crab pulsar} using IRFs with no correction factors applied produces fluxes not far from the desired value, albeit with distortions in the spectrum.

  \begin{figure}
    \resizebox{\hsize}{!}{\includegraphics{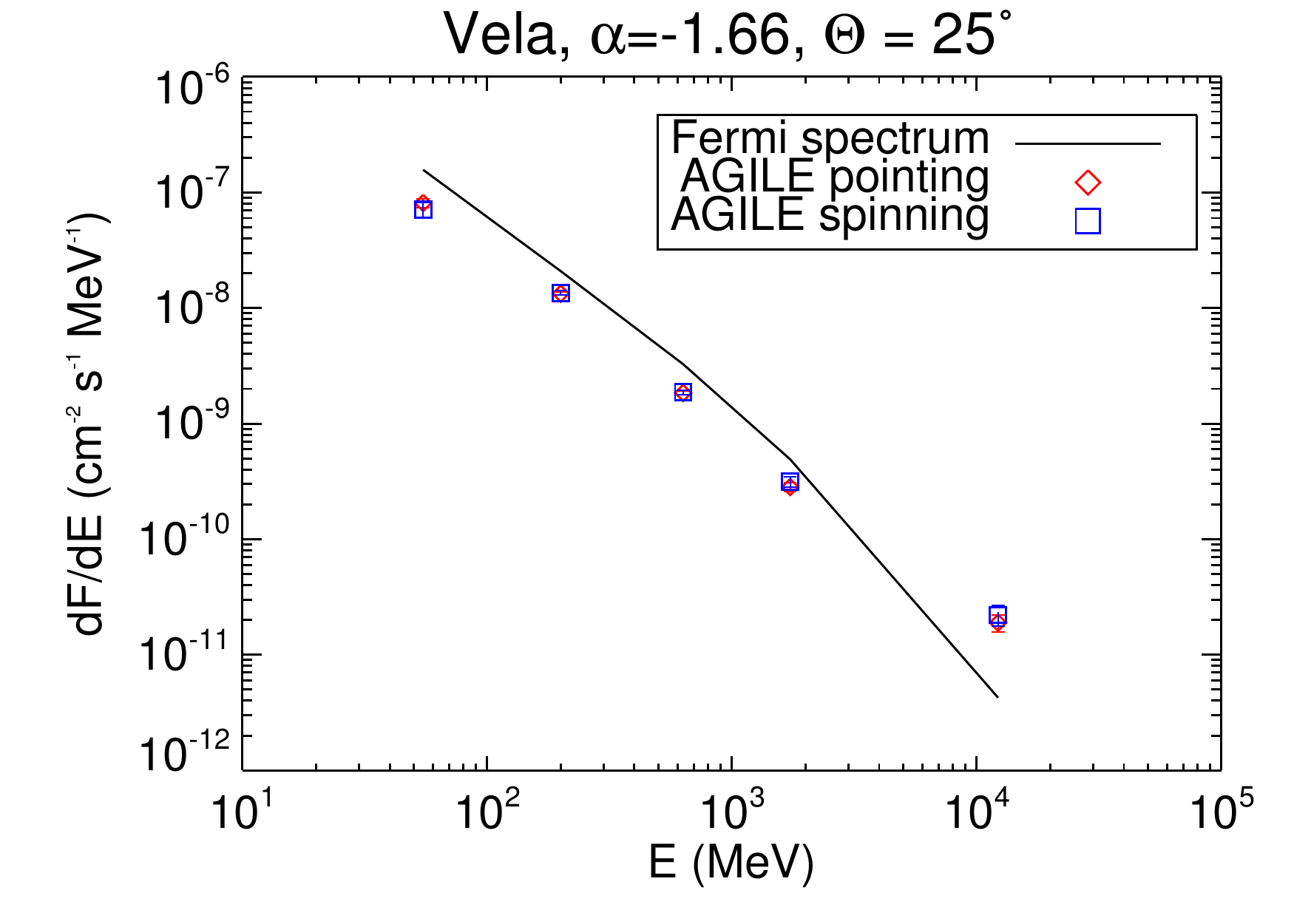}}
    \caption{Fluxes of the source at the position of the \object{Vela pulsar} found using the new effective area calculation (I0023) and the new PSF (I0023) for long integrations in pointing (red diamonds) and spinning (blue squares) mode. The black curve represents the flux and spectrum listed in 1FGL. No curve fitting was performed.
    }
    \label{scaling3}
  \end{figure}

  \begin{figure}
    \resizebox{\hsize}{!}{\includegraphics{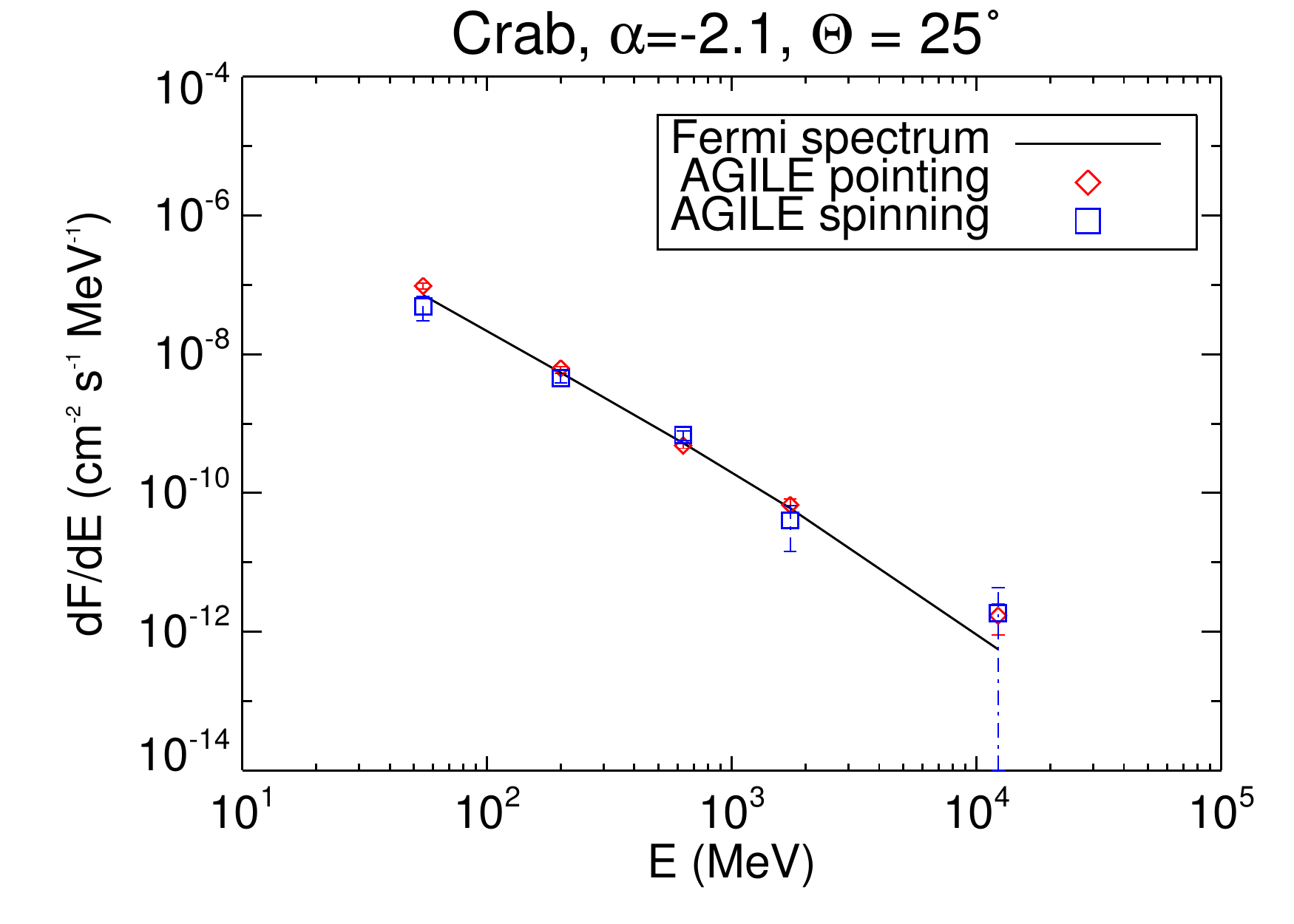}}
    \caption{Fluxes of the source at the position of the \object{Crab pulsar} found using the new effective area calculation (I0023) and the new PSF (I0023) for long integrations in pointing (red diamonds) and spinning (blue squares) mode. The black curve represents the flux and spectrum listed in 1FGL. No curve fitting was performed.
}
    \label{crab-scaling3}
  \end{figure}

\subsection{Fluxes and spectra: a new routine for generating exposure }

As a result, we concluded that scaling factors alone were unable to correct for the flux and spectra simultaneously for sources with both hard and soft spectra. We have revised the exposure generation routines to use the
true effective area formula in Eq. \ref{trueaeffeq}. We compare the results to the 1FGL spectra of Vela in Fig. \ref{scaling3} and the Crab in Fig. \ref{crab-scaling3}. In both cases, the AGILE analysis software assumes an unbroken power law with a
single spectral index and is therefore unable to model the exponential cutoff
above 2.9 GeV in the case of Vela and 5.8 GeV in the case of the Crab. Also, because 1FGL and the AGILE observations cover slightly different epochs, the Crab
flux and spectrum may be affected by variability (\citeads{2011Sci...331..736T}; \citeads{2011Sci...331..739A}).

\subsection{Point spread function}

\begin{figure}
    \resizebox{\hsize}{!}{\includegraphics{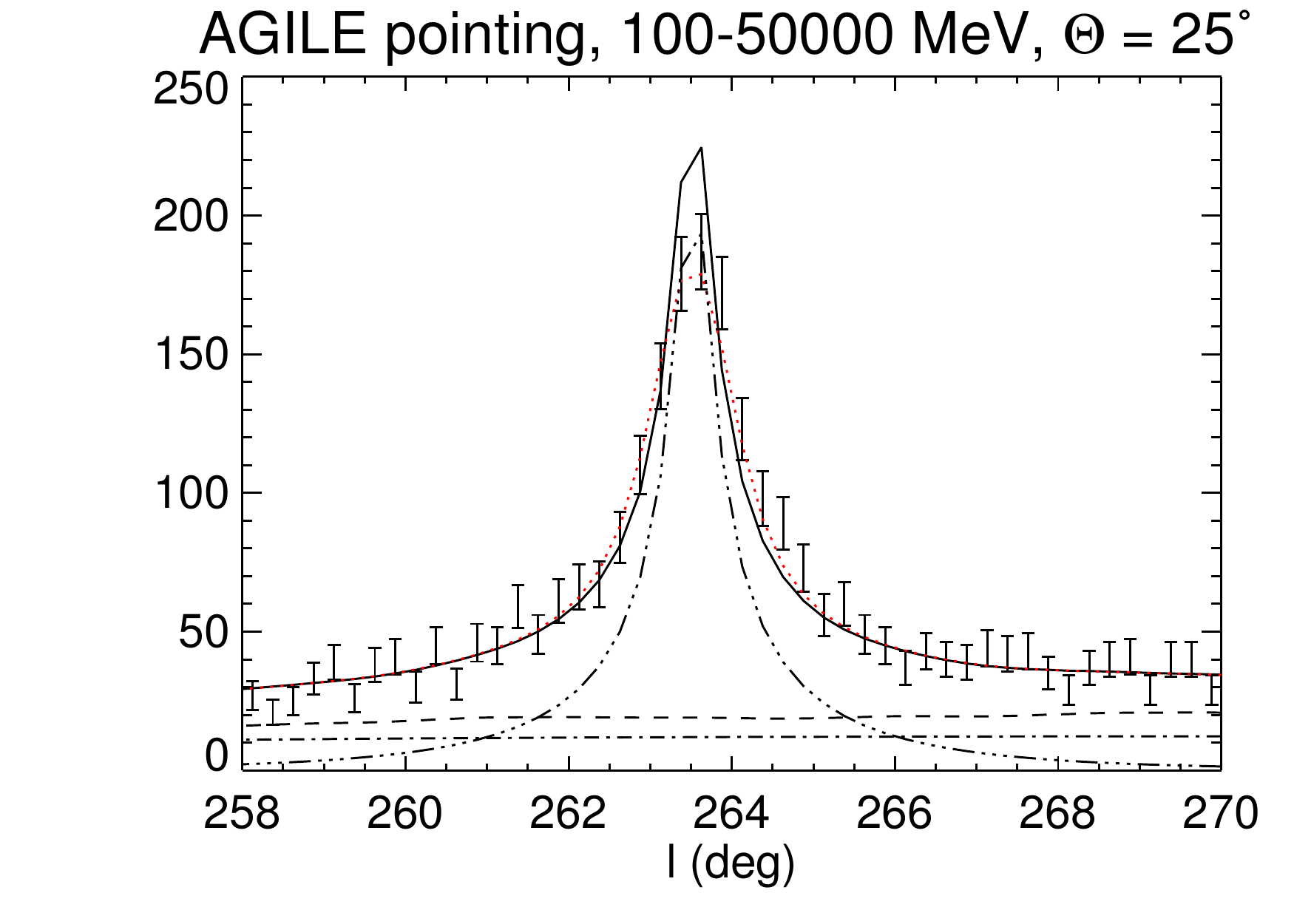}}
    \resizebox{\hsize}{!}{\includegraphics{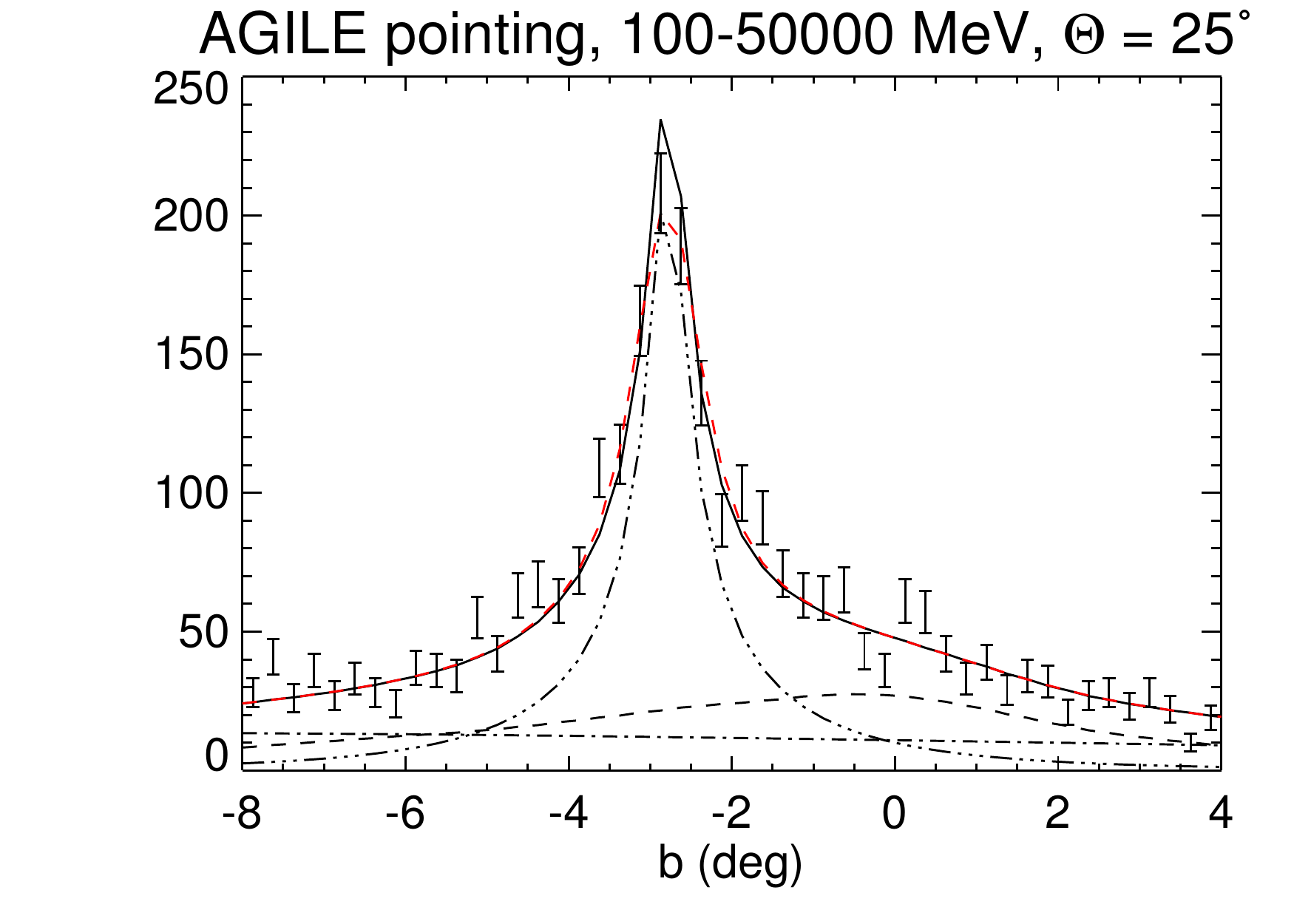}}
    \caption{Observed counts vs. model with PSF at the Vela source, $E > 100~$MeV, pointing mode.  The error bars are Poisson errors around the sum of the counts within $10^{\circ} \times 0.25^{\circ}$ slices in galactic longitude (top) and galactic latitude (bottom). The data are compared to a model (solid curve) composed of an isotropic component (dash-dot), a galactic diffuse component (dash), and a point source (dash-dot-dot-dot) with reduced $\chi_{\lambda}^2 = 1.41$ in longitude and 1.77 in latitude with 38 degrees of freedom. The composite PSF has a spectral index $\alpha = -1.66$, weighted by effective area, spectrum, and EDP. Smoothing the model with a Gaussian (red dotted) yields $\sigma=0.31$ and reduced $\chi_{\lambda}^2 = 0.81$ with 37 degrees of freedom  in longitude and $\sigma=0.22$ and reduced $\chi_{\lambda}^2 = 1.62$ in latitude, yielding likelihood ratios $\sqrt{TS} = 4.9$ and 2.7 respectively.}
    \label{psf1}
  \end{figure}

\begin{figure}
    \resizebox{\hsize}{!}{\includegraphics{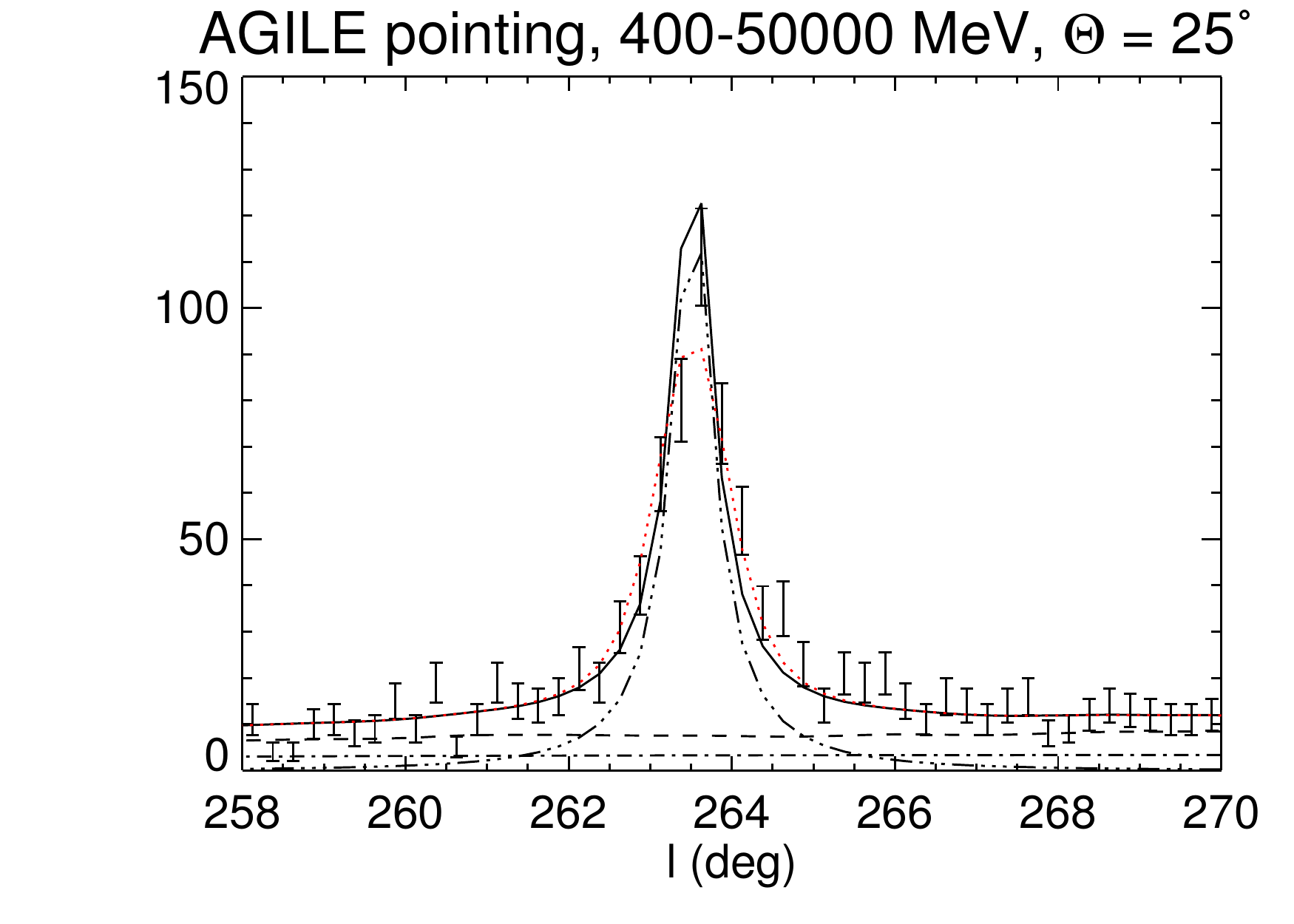}}
    \resizebox{\hsize}{!}{\includegraphics{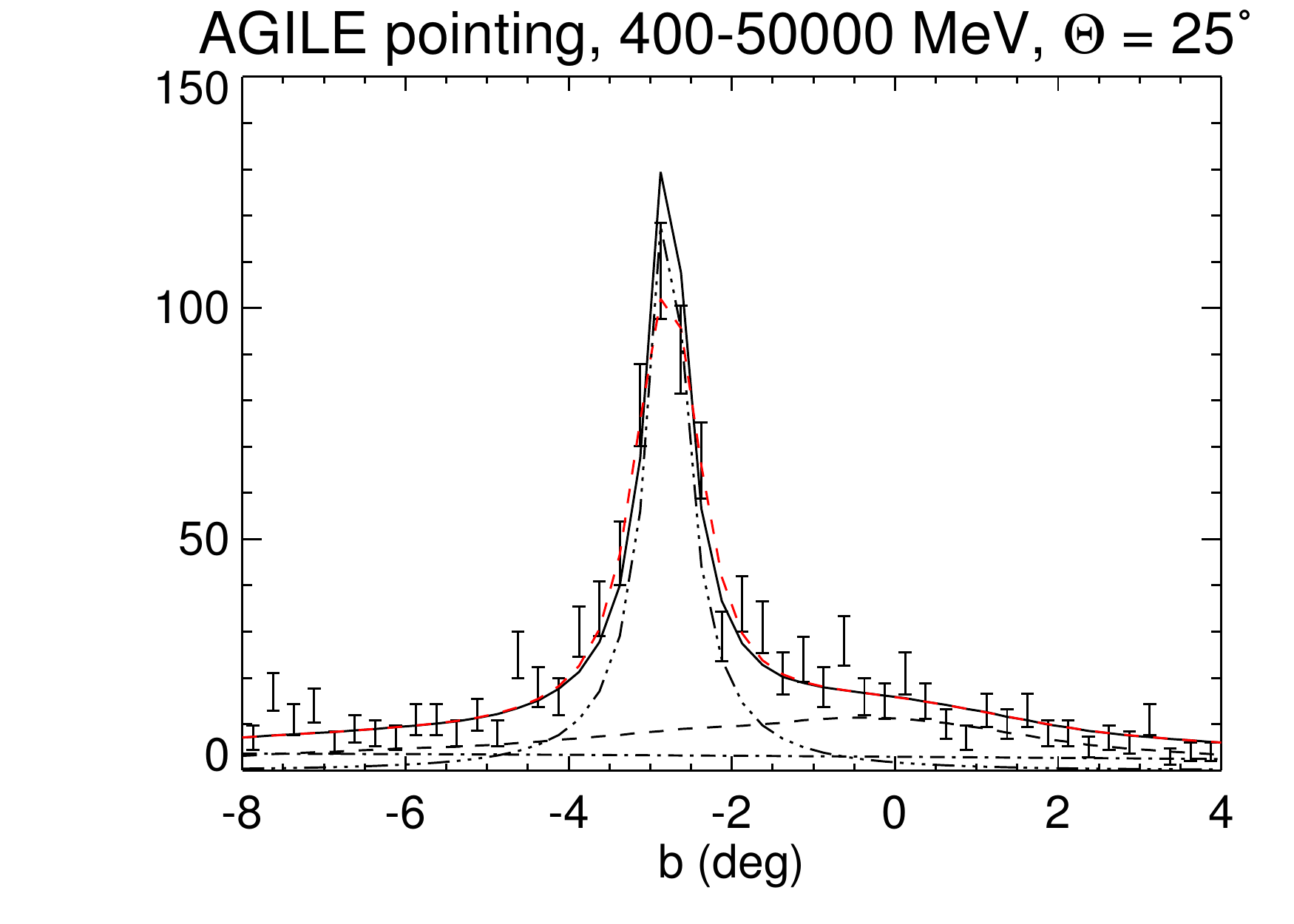}}
    \caption{Observed counts vs. model with PSF at the Vela source, $E > 400~$MeV, pointing mode. The error bars are Poisson errors around the sum of the counts within $10^{\circ} \times 0.25^{\circ}$ slices in galactic longitude (top) and galactic latitude (bottom). The data are compared to a model (solid curve) composed of an isotropic component (dash-dot), a galactic diffuse component (dash), and a point source (dash-dot-dot-dot) with reduced $\chi_{\lambda}^2 = 1.75$ in longitude and 1.68 in latitude with 38 degrees of freedom. The composite PSF has a spectral index $\alpha = -1.66$, weighted by effective area, spectrum, and EDP. Smoothing the model with a Gaussian (red dotted) yields $\sigma=0.30$ and reduced $\chi_{\lambda}^2 = 1.27$ with 37 degrees of freedom  in longitude and $\sigma=0.23$ and reduced $\chi_{\lambda}^2 = 1.41$ in latitude, yielding likelihood ratios $\sqrt{TS} = 4.4$ and 3.4 respectively.}
    \label{fig:psf2}
  \end{figure}

\begin{figure}
    \resizebox{\hsize}{!}{\includegraphics{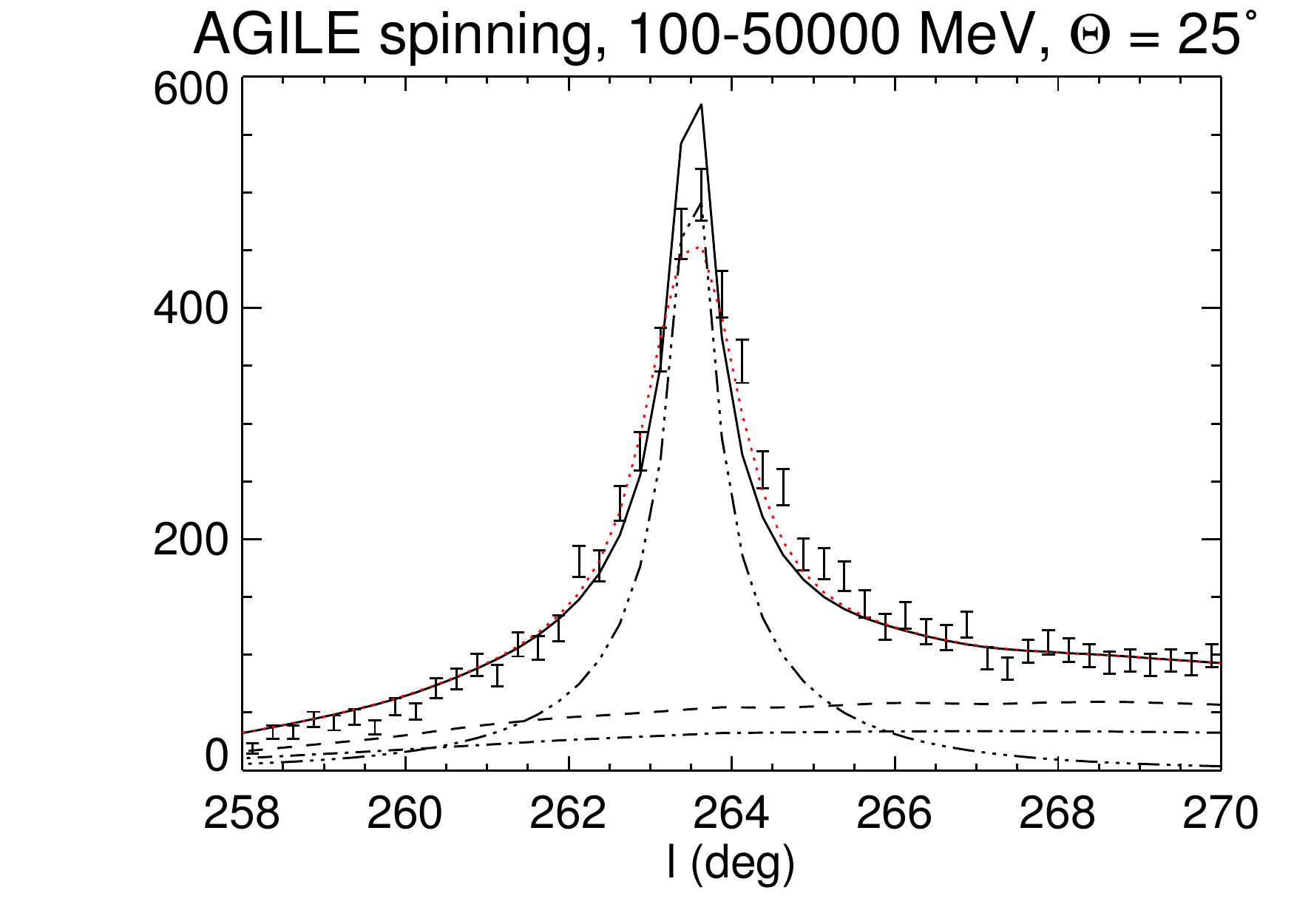}}
    \resizebox{\hsize}{!}{\includegraphics{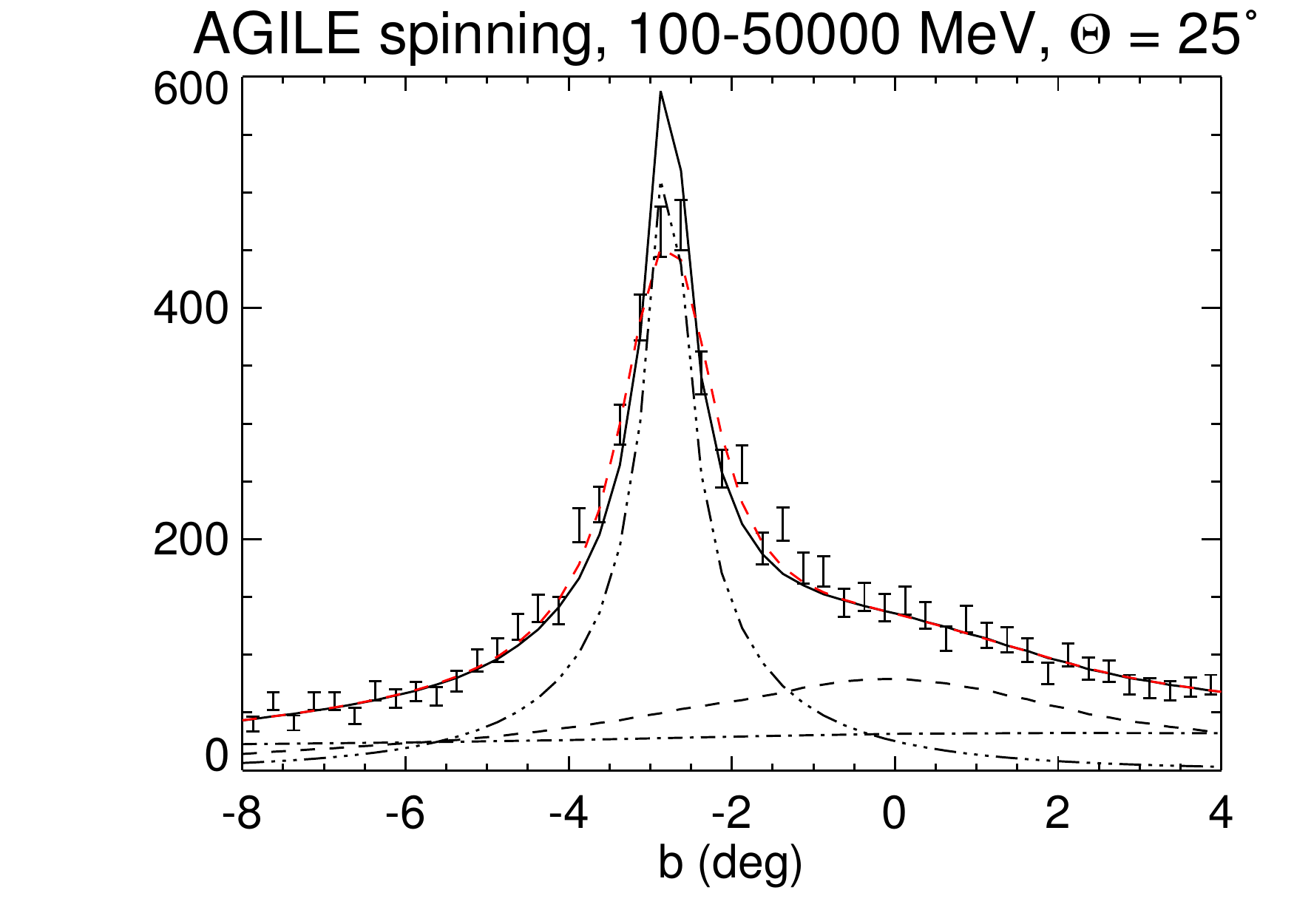}}
	    \caption{Observed counts vs. model with PSF at the Vela source, $E > 100~$MeV, spinning mode. The error bars are Poisson errors around the sum of the counts within $10^{\circ} \times 0.25^{\circ}$ slices in galactic longitude (top) and galactic latitude (bottom). The data are compared to a model (solid curve) composed of an isotropic component (dash-dot), a galactic diffuse component (dash), and a point source (dash-dot-dot-dot) with reduced $\chi_{\lambda}^2 = 3.42$ in longitude and 2.69 in latitude with 38 degrees of freedom. The composite PSF has a spectral index $\alpha = -1.66$, weighted by effective area, spectrum, and EDP. Smoothing the model with a Gaussian (red dotted) yields $\sigma=0.33$ and reduced $\chi_{\lambda}^2 = 1.96$ with 37 degrees of freedom  in longitude and $\sigma=0.33$ and reduced $\chi_{\lambda}^2 = 1.41$ in latitude, yielding likelihood ratios $\sqrt{TS} = 7.6$ and 7.1 respectively.}
    \label{psf3}
  \end{figure}

\begin{figure}
    \resizebox{\hsize}{!}{\includegraphics{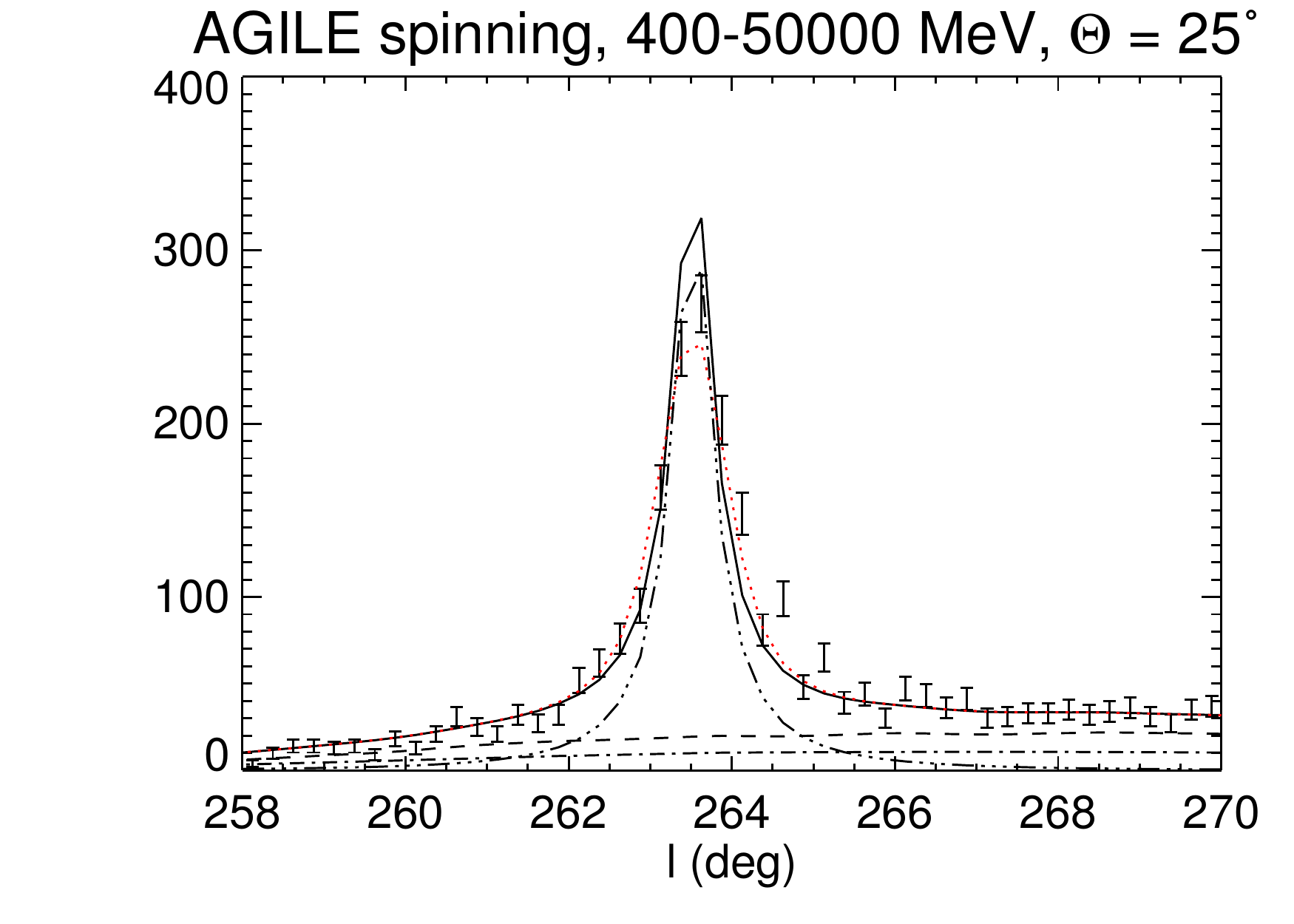}}
    \resizebox{\hsize}{!}{\includegraphics{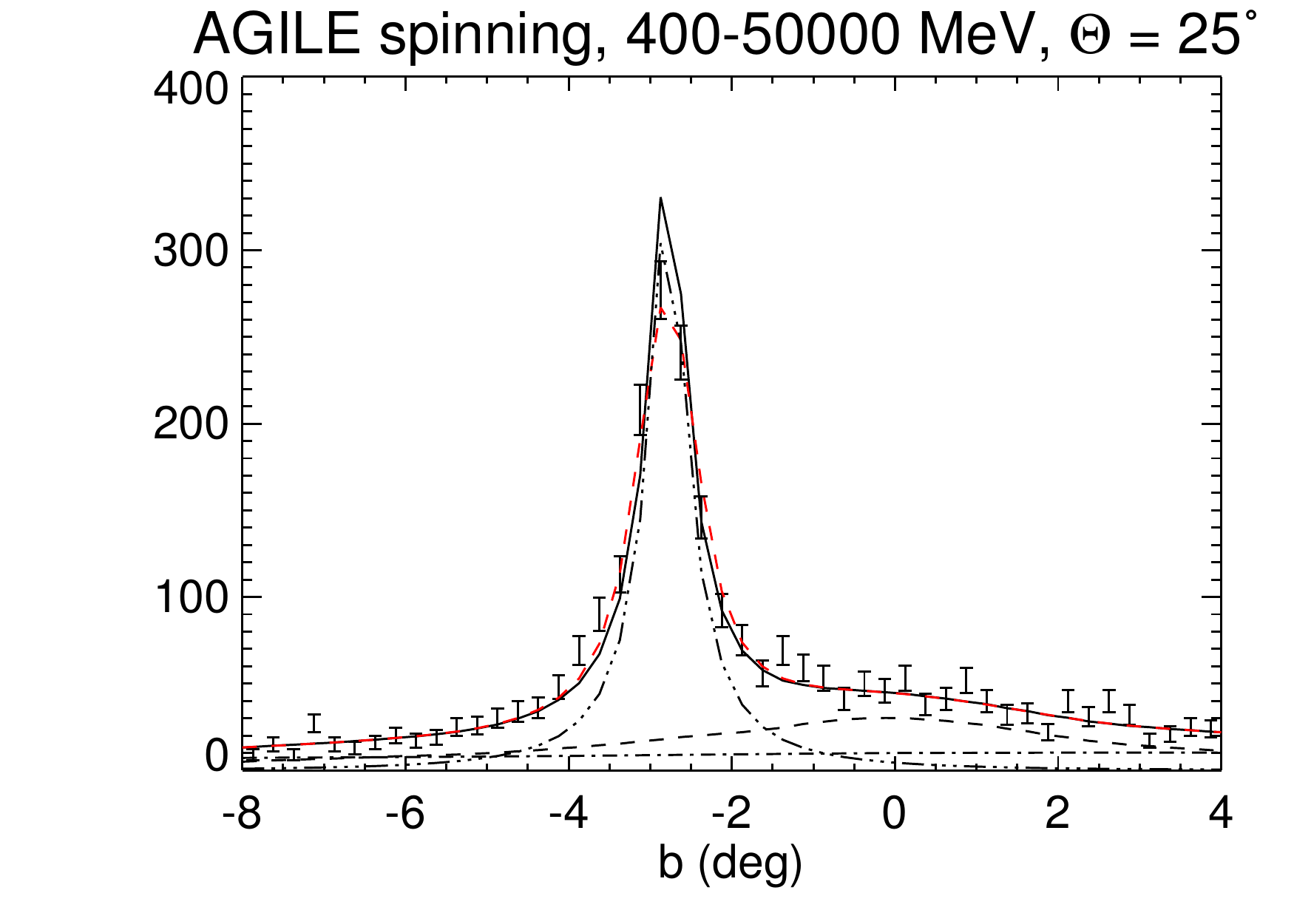}}
                                                                                                                                                                                                                                                                \caption{Observed counts vs. model with PSF at the Vela source, $E > 400~$MeV, spinning mode. The error bars are Poisson errors around the sum of the counts within $10^{\circ} \times 0.25^{\circ}$ slices in galactic longitude (top) and galactic latitude (bottom). The data are compared to a model (solid curve) composed of an isotropic component (dash-dot), a galactic diffuse component (dash), and a point source (dash-dot-dot-dot) with reduced $\chi_{\lambda}^2 = 2.82$ in longitude and 1.96 in latitude with 38 degrees of freedom. The composite PSF has a spectral index $\alpha = -1.66$, weighted by effective area, spectrum, and EDP. Smoothing the model with a Gaussian (red dotted) yields $\sigma=0.27$ and reduced $\chi_{\lambda}^2 = 1.75$ with 37 degrees of freedom  in longitude and $\sigma=0.21$ and reduced $\chi_{\lambda}^2 = 1.35$ in latitude, yielding likelihood ratios $\sqrt{TS} = 6.5$ and 4.9 respectively.}
   \label{psf4}
  \end{figure}

The PSFs as calculated in Eq. \ref{psfeq} were compared to the count maps
generated by the long integrations in pointing and spinning mode for all three pulsars, Vela, Crab, and Geminga, both as a function
of energy bin and for the full energy range from 100~MeV to 50~GeV.
The PSFs show
varying levels of agreement with the data.  Examples are shown in Figs. \ref{psf1}, \ref{fig:psf2}, \ref{psf3} and \ref{psf4}. In each of these figures, the number of counts were integrated within $10^{\circ} \times 0.25^{\circ}$ slices in galactic longitude and galactic latitude and compared to a model comprising an isotropic component, a galactic diffuse component, and a point source component  (see Eq.~\ref{psfeq}). The coefficients of the components were determined using the AGILE analysis software.

To estimate the goodness of fit, we calculated the maximum likelihood ratio statistic \citepads{1984NIMPR.221..437B}, 
\begin{equation}
\chi_{\lambda}^2 = 2 \sum_{i=1}^{N}[ M_i - C_i + C_i \ln(\frac{C_i}{M_i})],
\label{eq_ml}
\end{equation}
where $C_i$ is the number of counts and $M_i$ the number predicted by the model in each $10^{\circ}\times 0.25^{\circ}$ slice. The reduced $\chi_{\lambda}^2 $ is found by dividing by the number of degrees of freedom, which in this case is 38 (43 slices $- $ 3 free parameters).

\begin{figure}
  \centering
    \resizebox{\hsize}{!}{\includegraphics{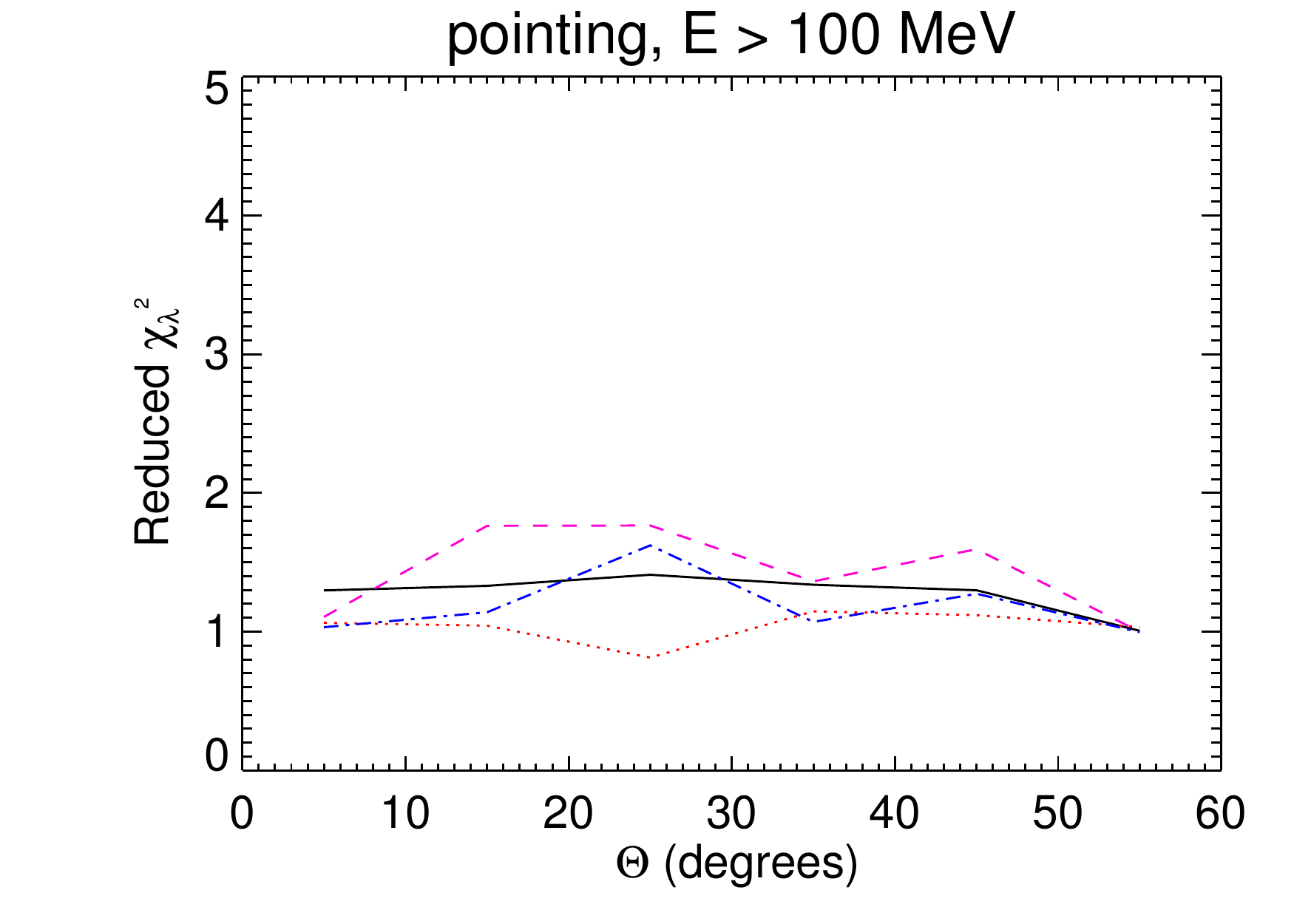}}
    \resizebox{\hsize}{!}{\includegraphics{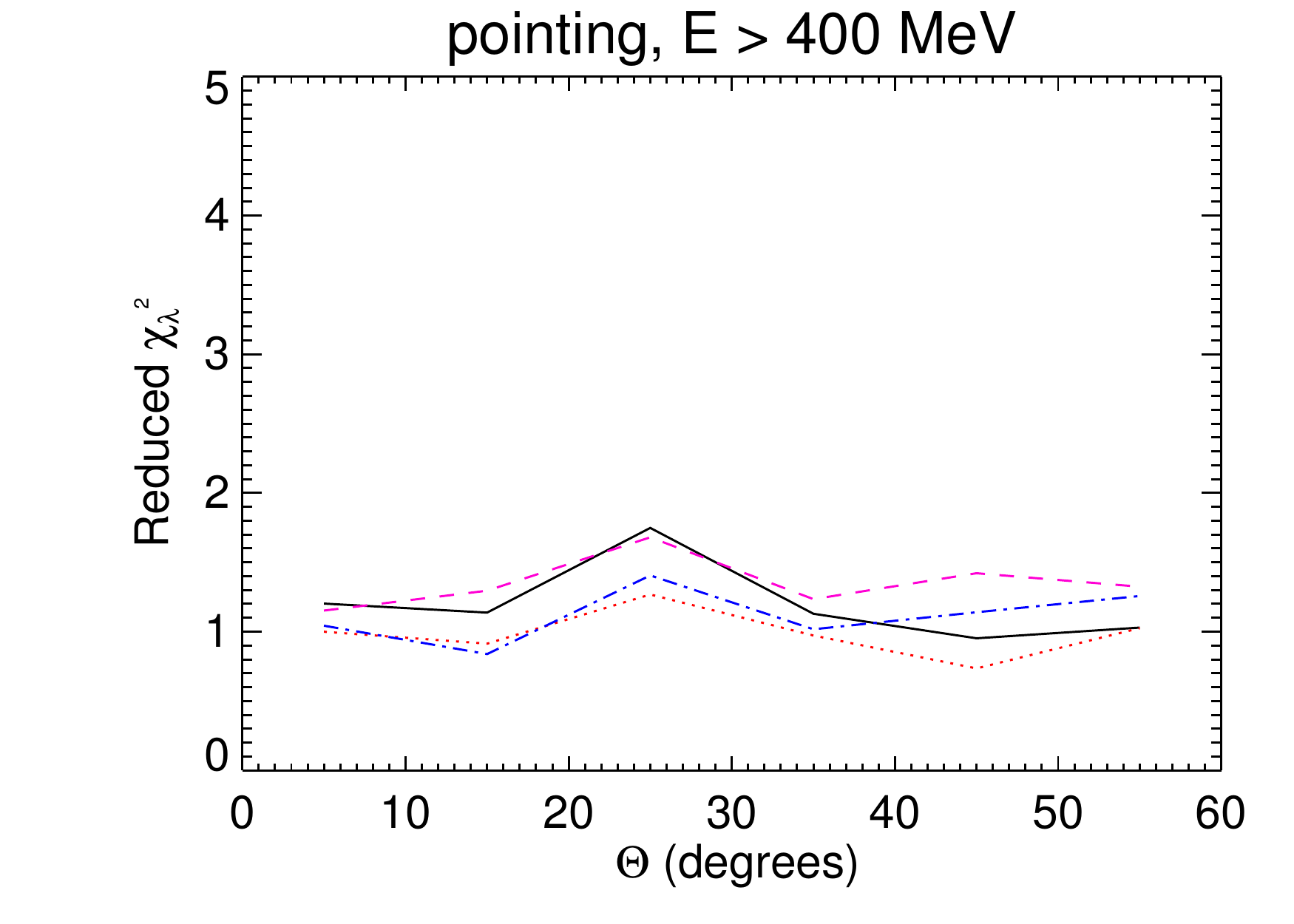}}
                                                                                                                                                                                                                                                                \caption{Reduced $\chi_{\lambda}^2$ for AGILE observations of $E > 100$~ MeV (top) and $E > 400$~ MeV (bottom) in pointing mode. Galactic longitude slices with (red dotted) and without (solid black) Gaussian smoothing of the model; Galactic latitude slices with (blue dot-dash) and without (magenta dashed) Gaussian smoothing. 38 degrees of freedom without and 37 with smoothing. The unsmoothed model shows good agreement and the fit is not significantly improved by Gaussian smoothing.}

    \label{fig:chi_point}
  \end{figure}

\begin{figure}
  \centering
    \resizebox{\hsize}{!}{\includegraphics{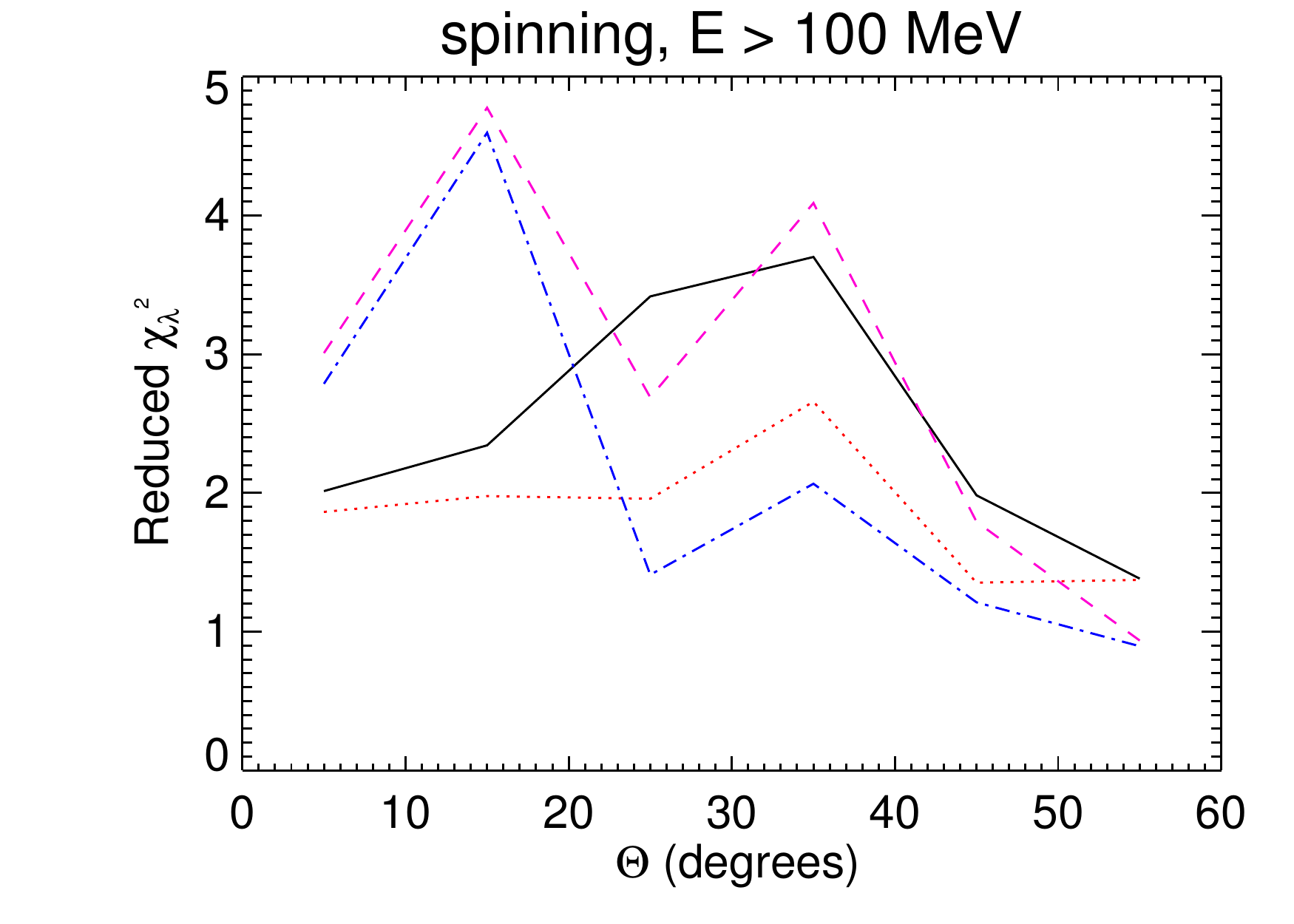}}
    \resizebox{\hsize}{!}{\includegraphics{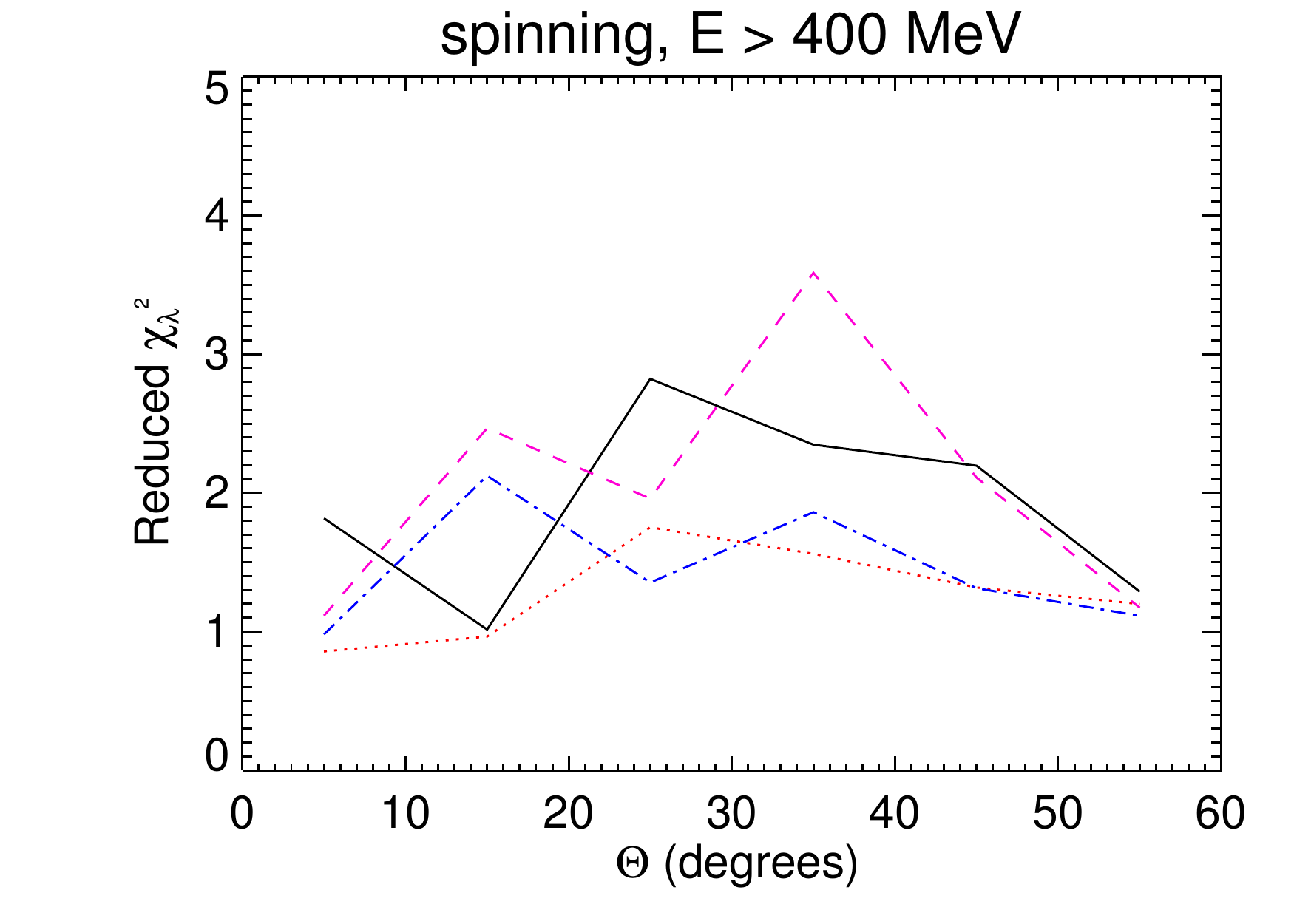}}
                                                                                                                                                                                                                                                                \caption{Reduced $\chi_{\lambda}^2$ for AGILE observations of $E > 100$~ MeV (top) and $E > 400$~ MeV (bottom) in spinning mode. Galactic longitude slices with (red dotted) and without (solid black) Gaussian smoothing of the model; Galactic latitude slices with (blue dot-dash) and without (magenta dashed) Gaussian smoothing. 38 degrees of freedom without and 37 with smoothing. In many cases, Gaussian smoothing significantly improves the goodness of fit.}

    \label{fig:chi_spin}
  \end{figure}

In some cases, the real PSF appears to be broader than the model predicts, particularly in spinning mode. One possible source of this broadening is systematic error in the measurement of the spacecraft orientation. For each AGILE observation in both pointing and spinning mode, we smoothed the two-dimensional model with a simple Gaussian and found the Gaussian width $\sigma$ which minimized $\chi_{\lambda}^2$. The difference  $\sqrt{TS}=$ unreduced  $\chi_{\lambda}^2(\sigma) -$ unreduced $\chi_{\lambda}^2(0)$ should be distributed as $\chi^2$ with one degree of freedom and therefore be statistically significant when it is greater than 5. Best fit Gaussian smoothed model PSFs are shown in Figs. \ref{psf1}, \ref{fig:psf2}, \ref{psf3} and \ref{psf4}. Figs. \ref{fig:chi_point} and \ref{fig:chi_spin} show the reduced $\chi_{\lambda}^2$ with and without Gaussian smoothing. 

\begin{figure}
  \centering
    \resizebox{\hsize}{!}{\includegraphics{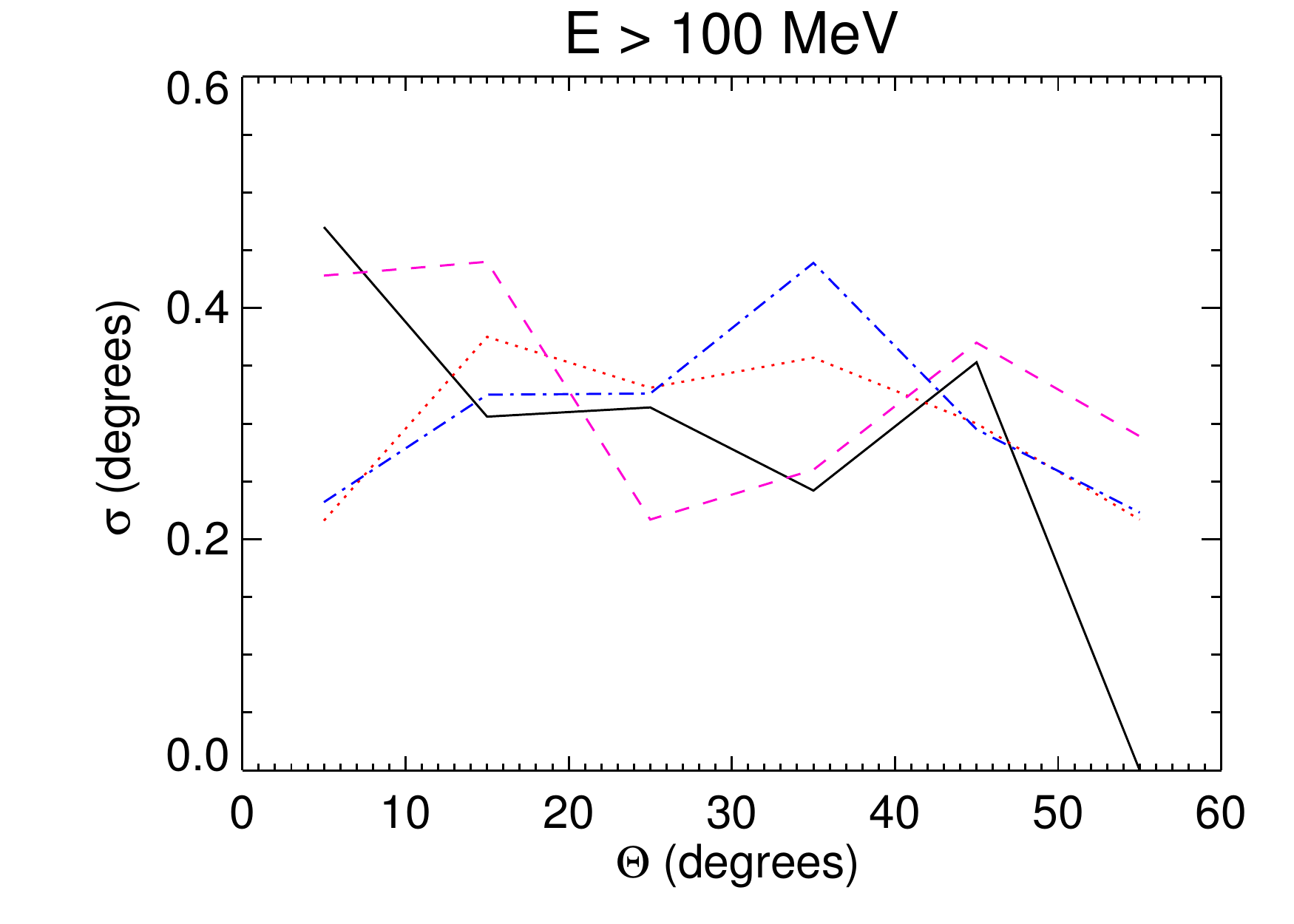}}
    \resizebox{\hsize}{!}{\includegraphics{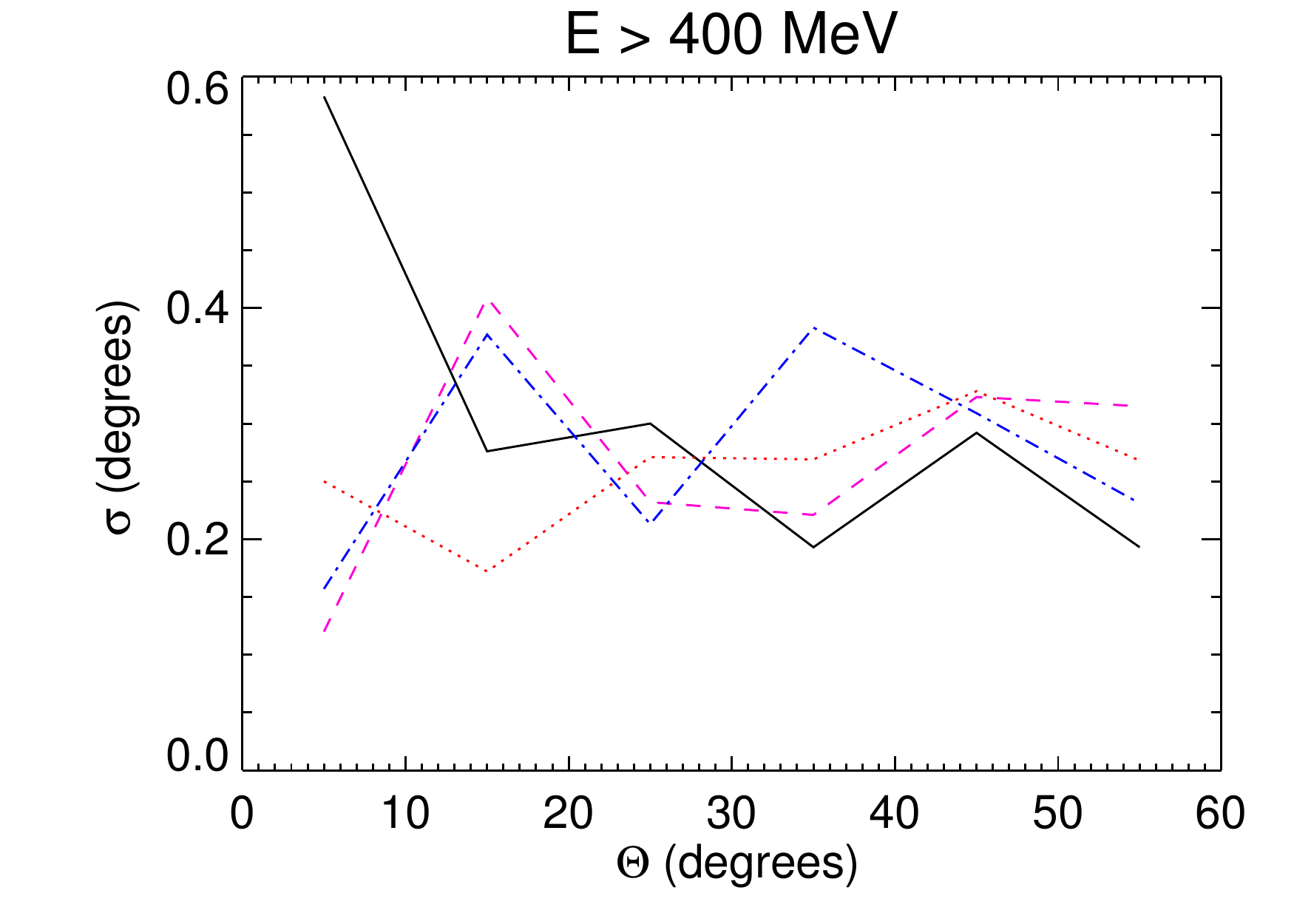}}
                                                                                                                                                                                                                                                                \caption{Best-fit Gaussian smoothing width $\sigma$ for AGILE observations of $E > 100$~ MeV (top) and $E > 400$~ MeV (bottom). Galactic longitude slices in pointing (solid black) and spinning (red dotted) mode; Galactic latitude slices in pointing (magenta dashed) and spinning (blue dot-dash) mode. The widths are roughly consistent with $\approx 0.3^{\circ}$.}

    \label{fig:sigma}
  \end{figure}

\begin{figure}
  \centering
    \resizebox{\hsize}{!}{\includegraphics{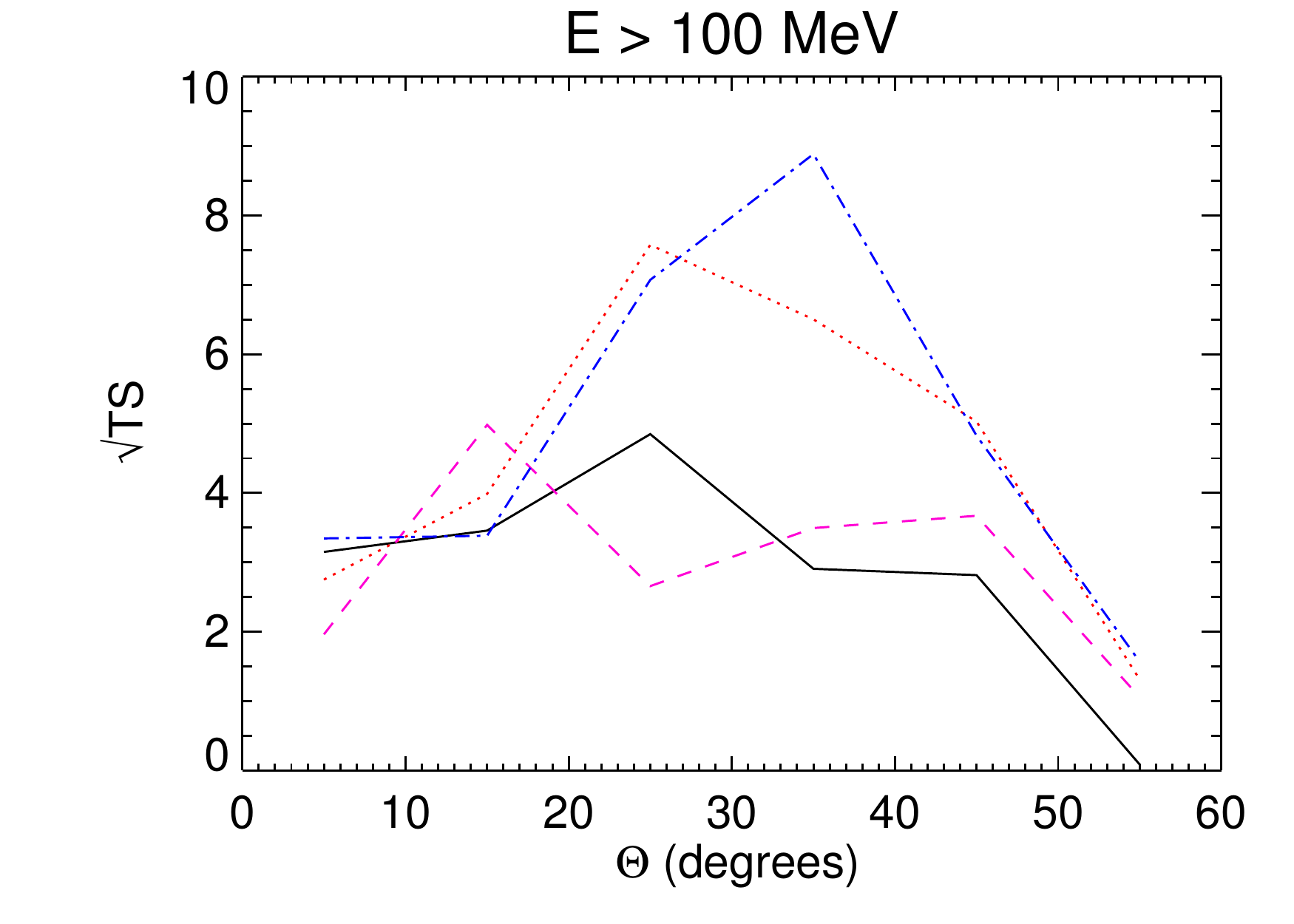}}
    \resizebox{\hsize}{!}{\includegraphics{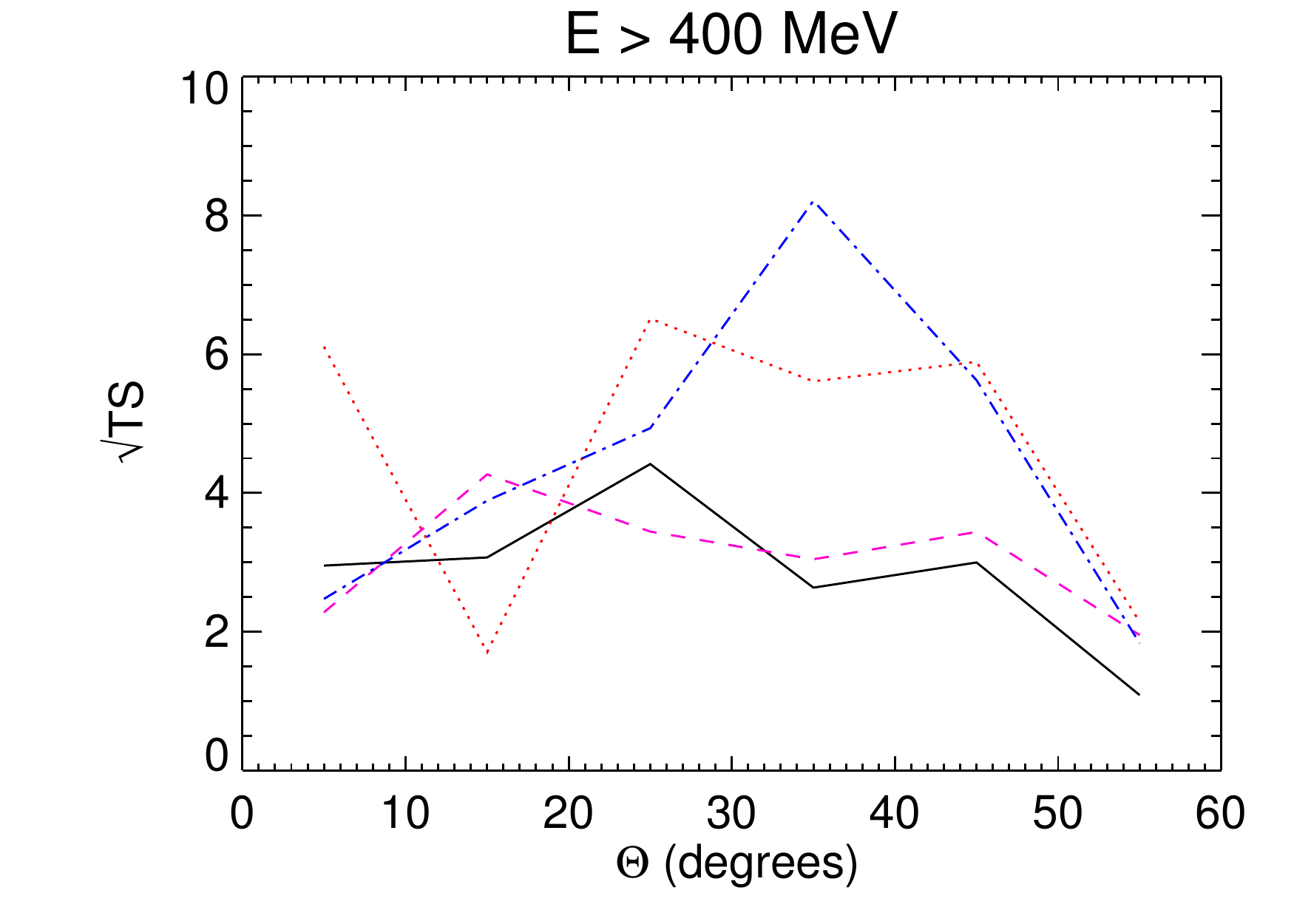}}
                                                                                                                                                                                                                                                                \caption{Significance of improvement $\sqrt{TS}$ due to additional parameter $\sigma$ for AGILE observations of $E > 100$~ MeV (top) and $E > 400$~ MeV (bottom). Galactic longitude slices in pointing (solid black) and spinning (red dotted) mode; Galactic latitude slices in pointing (magenta dashed) and spinning (blue dot-dash) mode. $\sqrt{TS}~>~5$ only for observations in spinning mode.}

    \label{fig:sqrtts}
  \end{figure}

Fig. \ref{fig:sigma} shows the best fit $\sigma$ as a function of $\Theta$. The values of $\sigma$ are roughly consistent with $\approx 0.3^{\circ}$. However, in Fig. \ref{fig:sqrtts}, we see that $\sqrt{TS}$ shows a statistically significant improvement only in the case of the observations in spinning mode. These results are consistent with the hypothesis of a systematic error in the measurement of the spacecraft orientation in spinning mode.


A refined attitude reconstruction method using Kalman filtering techniques,
optimized for the AGILE spinning observation mode, has been recently developed by the Compagnia Generale per lo Spazio (CGS), in joint collaboration with the ASDC. CGS is the prime industrial contractor of the AGILE mission, being in charge of design development and integration of the complete satellite. Star Sensor data in spinning mode are noisier, and present short gaps due to occasional blinding. The new attitude reconstruction improves the efficiency and the quality of the attitude measurement. A new analysis of in-flight spinning data reprocessed with the new attitude reconstruction is in progress at ASDC, and the results will be presented elsewhere.

\section{Conclusions}

The on-ground background rejection filters used by AGILE-GRID have
been optimized a number of times to increase the effective area while maintaining
a reasonable level of instrumental and cosmic-ray background. To validate and
keep pace with these changes, the monoenergetic PSFs and EDPs produced by Monte Carlo simulations and validated by pre-launch tests were compared to in-flight data. 

The effective area calculations in narrow and wide reconstructed energy bands show extreme sensitivity to the assumed spectral index due to the large energy dispersion. As a result, for day-to-day analysis, correction factors were calculated and introduced into the effective area matrices as a substitute for the full energy dispersion calculation. 

These correction factors produced valid results only for a limited range of
source spectra. A new version of the analysis software, soon to be released
by the ASDC, properly takes into account the energy dispersion when calculating
the energy-dependent effective area. The software may now be used to calculate the spectral index through simultaneous analysis of the data divided into energy intervals. By comparing the calculated index to the index initially assumed to generate the exposure files and PSFs and iterating, the true flux and spectral index of the source may then be found. Strong deviations from power-law spectral behavior are not implemented and may lead to distortions, particularly at low and high energies where a large portion of the flux may come from outside the nominal energy bins.

The in-flight PSFs for real sources in pointing mode agree with those predicted by the Monte Carlo simulations, while those in spinning mode differ significantly. This effect is probably due to systematic error in the Star Sensor measurement of the spacecraft orientation in spinning mode.  A new optimized attitude reconstruction method currently under testing at ASDC should be able to correct this systematic error, which broadens the PSF by $\approx 0.3^{\circ}$ for spinning mode observations.
 
AGILE and Fermi have different pointing strategies and are sensitive to
variability on different timescales. In addition, at any given time AGILE and Fermi pointed toward different areas on the sky. AGILE-GRID therefore remains a completely complementary instrument 
for the detection of rapid transient phenomena.

\begin{acknowledgements}
We would like to thank the Istituto Nazionale di Astrofisica, the Agenzia Spaziale Italiana, the Consorzio Interuniversitario per la Fisica Spaziale, and the Istituto Nazionale di Fisica Nucleare for their generous support of the AGILE mission and this research, including ASI contracts n. I/042/10/1 and I/028/12/0. We would also like to thank the journal referee, whose comments helped to substantially improve this paper.
\end{acknowledgements}


\bibliographystyle{aa}

\bibliography{extras,calib_final}

\end{document}